\documentclass[review,11pt]{elsarticle}

\usepackage{lineno,hyperref}
\usepackage{bm}
\usepackage{algorithm}
\usepackage{algorithmicx}
\usepackage[noend]{algpseudocode}
\let\Algorithm\algorithm
\usepackage{caption}
\usepackage{hyperref}
\usepackage[utf8]{inputenc}
\usepackage{lscape}
\usepackage{tikz}
\usepackage{float}
\usepackage{graphicx}
\usepackage{listings}%
\usepackage{amsmath,amsfonts}
\usepackage{mathtools}
\usepackage{wrapfig}
\usepackage{amssymb}
\usepackage{alltt}%
\usepackage{siunitx}
\usepackage{setspace}
\usepackage{eqparbox}
\usepackage{booktabs}
\usepackage[margin=2cm]{geometry}

\renewcommand\algorithm[1][]{\Algorithm[#1]\setstretch{1.4}}

\usepackage[labelformat=simple]{subcaption} 

\DeclareCaptionLabelFormat{subcaptionlabel}{\normalfont(\textbf{#2}\normalfont)}
\journal{arXiv}

\biboptions{sort&compress}

\begin{document}
\nolinenumbers
\begin{frontmatter}

\title{Solving the Discretised Neutron Diffusion Equations using Neural Networks} 

\author[<address link1>]{Toby R.F. Phillips}
\author[<address link1>]{Claire E. Heaney}
\author[<address link1>]{Boyang Chen}
\author[<address link2>]{Andrew G. Buchan}
\author[<address link1>]{Christopher C. Pain}
\address[<address link1>]{{Applied Modelling and Computation Group, Department of Earth Science and Engineering}, {Imperial College London},
{{London}, {SW7 2AZ} {United Kingdom}}}
\address[<address link2>]{{School of Engineering and Materials Science}, {Queen Mary University of London},
{{London}, {E1 4NS} {United Kingdom}}}

\begin{abstract}

This paper presents a new approach which uses the tools within Artificial Intelligence (AI) software libraries as an alternative way of solving partial differential equations (PDEs) that have been discretised using standard numerical methods. 
In particular, we describe how to represent numerical discretisations arising from the finite volume and finite element methods by pre-determining the weights of convolutional layers within a neural network. As the weights are defined by the discretisation scheme, no training of the network is required and the solutions obtained are identical (accounting for solver tolerances) to those obtained with standard codes often written in Fortran or C++.  We also explain how to implement the Jacobi method and a multigrid solver using the functions available in AI libraries. For the latter, we use a U-Net architecture which is able to represent a sawtooth multigrid method. A benefit of using AI libraries in this way is that  one can exploit their power and their built-in technologies. For example, their executions are already optimised for different computer architectures, whether it be CPUs, GPUs or new-generation AI processors. 

In this article, we apply the proposed approach to eigenvalue problems in reactor physics where neutron transport is described by diffusion theory. For a fuel assembly benchmark, we demonstrate that the solution obtained from our new approach is the same (accounting for solver tolerances) as that obtained from the same discretisation coded in a standard way using Fortran. We then proceed to solve a reactor core benchmark using the new approach. 

\end{abstract}

\begin{keyword}
Numerical solution of partial differential equations; Finite Difference Method; Finite Volume Methods; Convolutional Neural Network; Multigrid Solver; U-Net; Neutron Diffusion Equation; Reactor Physics
\end{keyword}

\end{frontmatter}


\section{Introduction}
Development of new computational hardware brings with it the challenge of adapting code in order for it to be deployed successfully on these new architectures. In the field of Artificial Intelligence (AI), this challenge has largely been met by writers of and contributors to widely used AI libraries (for example TensorFlow~\citep{tensorflow2015} and PyTorch~\citep{pytorch}). In these libraries, code relating to the architecture has been abstracted away so that users can concentrate on the algorithm they wish to implement without having to think about or understand the code relating to the computer architecture. As a result, the user has only to make minimal changes to their code in order to run on Central Processing Units (CPUs) or Graphical Processing Units (GPUs) or even Tensor Processing Units (TPUs). In other fields, such as scientific computation, perhaps because the codes and libraries are less standard and more numerous, users have to expend much more effort to run their codes on new architectures. Porting code to clusters of CPUs is relatively straightforward nowadays, however the computational gains to be had by running on clusters are limited by memory access and data transfer. Although GPUs have demonstrated superior performance to CPUs, instructions for the GPU must be written in languages such as CUDA or OpenCL that are unfamiliar to many working in scientific computation. This additional coding task has hindered the take-up of GPUs, although there are examples of this having been done successfully, for example, in computational fluid dynamics~\citep{Vermiere2017}, for acoustic waves~\citep{Chan2016} and, in radiation transport, for a Monte Carlo neutron transport code~\citep{Bergmann2017} and for eigenvalue problems~\citep{Slaybaugh2018}. With CUDA and OpenCL, GPUs have been used to accelerate generation of finite element matrices for unstructured meshes~\citep{Cecka2011,Mossaiby2012,Dziekonski2013,Sanfui2020} and for discontinuous Galerkin methods~\citep{Modave2017}.
Recently, new types of processors have been unveiled, which have been designed specifically for tasks associated with AI such as matrix multiplication and vector operations. These processors are therefore also suited to the linear algebra calculations that arise in the area of scientific computation~\citep{Lewis2022}. Furthermore, these new processors are designed to be more energy efficient than CPUs or GPUs, with hundreds of thousands of cores on a single chip, making it ideal hardware for researchers to run computationally demanding problems in an energy-efficient manner. AI libraries are already up-and-running on these so-called AI processors, which include TPUs of Google~\citep{Lewis2022}, Intelligence Processing Units (IPUs) of Graphcore~\citep{GraphCore_IPU} and CS-2 of Cerebras~\citep{cerebras}. In order to exploit the speed of GPUs or AI processors for scientific computations, this paper outlines a method of formulating numerical discretisations in terms of operations or functions found in AI libraries, such as discrete convolutions. Writing discretisations in this way means that code can be deployed on whichever platform is available, whether it be CPUs, GPUs or the new AI processors, without having to make major modifications to the code.

Previous work that exploits the linear algebra capabilities of AI processors by using AI libraries to solve scientific problems includes applications in distributed Fourier Transforms~\citep{Lu2020,Lu2021}; Monte Carlo simulations for finance~\citep{Belletti2020}; many-body quantum physics~\citep{Morningstar2022}; and density functional theory \citep{Pederson2022}. We have found four examples of previous work that exploits operations associated with neural networks that can be found within AI libraries in order to solve scientific problems~\citep{Zhao2020,Wang2022,Chen2023,Phillips2022-progress}. \citet{Zhao2020} were the first to equate a finite difference discretisation of the Navier-Stokes equations with a convolutional neural network in which the weights were determined by the discretisation. For validation, they use a number of benchmark tests including lid-driven cavity flow and flow past a cylinder. \citet{Wang2022} present a similar idea to~\citet{Zhao2020}, again using TensorFlow to implement finite difference discretisations of CFD problems, however using TPUs rather than GPUs. They solve the variable-density Navier-Stokes equations and demonstrate good weak and strong scaling. \citet{Chen2023} implement both a finite difference and a finite element discretisation through convolutional neural networks in order to solve a number of CFD problems. They develop a method of solving the discretised systems based on a combination of a sawtooth multigrid method and the Jacobi method implemented as a U-Net~\citep{Ronneberger2015} (a convolutional neural network with a specific architecture). Using convolutional neural networks with pre-determined weights, \citet{Phillips2022-progress} implement an upwind finite volume discretisation and several finite element discretisations arising from a new convolutional finite element method (ConvFEM). The application they study, radiation transport, requires development of a 4D multigrid method, again, based on the U-Net. Researchers have previously noted the similarity between the multigrid method and the encoder-decoder type of neural networks, such as the U-Net~\citep{Ronneberger2015}. Consequently, the use of multigrid-inspired architectures for (trained) neural networks has been explored and been shown to enhance performance relative to conventional CNN architecture for applications in computer vision \citep{Ke2017,He2019} and in computational fluid dynamics (CFD)~\citep{Thuerey2020,Le2021}. 
Taking a different approach, \citet{Margenberg2022} use a (trained) neural network to produce solutions for the finer levels of a multigrid and standard CFD solvers to produce solutions at the coarser levels. 
By contrast to the previous examples of integrating the multigrid method with neural networks~\citep{Ke2017,He2019,Thuerey2020,Le2021,Margenberg2022}, \citet{Chen2023} implements a multigrid method using (untrained) neural networks with pre-determined weights to solve the PDEs on the coarse levels, and determine the residuals and provide Jacobi relaxation on finer levels. It is this method that is adopted in our current investigation.

In this paper we describe how to implement a finite volume discretisation of the neutron diffusion equation using a convolutional neural network whose weights are pre-determined by the particlar discretisation scheme. (In this case our finite volume discretisation is equivalent to a finite difference discretisation.) We also use this approach to implement a quadratic finite-element discretisation for the neutron diffusion equation. The Jacobi method and a sawtooth mutligrid method are used as solvers, implemented through standard operations found in AI software libraries. We demonstrate the approach using a fuel assembly benchmark, and compare the solution for the spatial variation of the neutron flux with that obtained from the same discretisation coded in a standard way using Fortran. We then proceed to solve a reactor core benchmark using the proposed method. The approach described in this article is a new and alternative way of harnessing AI technologies for forming solutions of governing PDEs. Ultimately solutions obtained through this new approach are identical to those obtained by standard codes, but the advantage of performing all operations through an AI library is that the code will run efficiently on all architectures. Furthermore, through the neural networks, the latest developments can be realised for methods such as sensitivities~\citep{cervi2022sensitivity} , uncertainty quantification~\citep{kabir2018neural,Tripathy2018} and data assimilation~\citep{Gong2022}. Although AI is becoming popular for nuclear engineering, it is often through surrogate modelling, which requires training a neural network. Some examples of current work include using physics-informed neural networks for point kinetics~\citep{SCHIASSI2022108833} and for non-smooth heterogeneous neutron diffusion problems~\cite{WANG2022109234}; surrogate models for transient analysis~\citep{FOAD2022109017}, eigenvalue problems~\citep{phillips2021} and digital twins~\citep{Gong2022}; and cross-section generation with neural networks~\citep{QIN2020107785}. The approach presented here is fundamentally different, however, providing an alternative way of exactly representing a given discretisation of a system of PDEs, whereas surrogate models provide an approximation of a discretised system of PDEs. Having formulated these discretisations in terms of neural networks, an obvious extension is to combine both untrained networks (i.e.~the networks with pre-determined weights as described here) and trained networks to form more efficient and powerful digital twins, as has previously been observed~\citep{Zhao2020,Beck2021,Chen2023}.

The sections of this paper are organised as follows. Section~\ref{methods} describes how a convolutional neural network can be uesd to express a finite volume discretisation, and how a neural network can be used to formulate Jacobi iterations and multigrid methods. Section~\ref{results} presents the three numerical examples using the neural network solver to resolve reactor physics eigenvalue problems, and comparisons are drawn against a standard finite volume method. Finally, Section~\ref{conclusion} completes the paper with a conclusion of its findings. 

\section{Methodology}\label{methods}

The first part of this section introduces the governing equations, their discretisation with the finite volume or control volume method and the Jacobi method for solving the resulting system. We then explain how the discretisation can be formulated using convolutional layers of a neural network with pre-defined weights. To solve the resulting system, we embed a Jacobi method within a multigrid method. Both Jacobi and multigrid methods are implemented within the neural network, and the latter is based on the U-Net architecture. Finally, an overview of the solution process is given for the case of multiple energy groups, including how the eigenvalue is determined. 
\subsection{Diffusion Equation}\label{diffusion_equation} 
The multi-group steady-state diffusion equation for criticality can be written as:
\begin{equation}\label{eq:diff-eig}
\begin{split} 
- \nabla\cdot (D_g \nabla \phi_g) + \Sigma^a_g \phi_g + \sum_{\substack{g^{'}= 1\\ g^{'}\neq g}}^{N_g}\Sigma^s_{g\rightarrow g^{'}}\phi_g &= \sum_{\substack{g^{'}= 1\\ g^{'}\neq g}}^{N_g}\Sigma^s_{g^{'}\rightarrow g}\phi_{g^{'}} + \lambda\,\chi_g\sum_{g^{'}=1}^{N_g} \nu_{g^{'}} \Sigma^f_{g^{'}} \phi_{g^{'}},
\\
& \forall g\in\{1,2,\ldots,N_g\},
 \end{split}
\end{equation}
where $\phi_g$ is the scalar flux of the neutron population, $\Sigma^a_g$ represents the absorption cross-section, $\Sigma^f_g$ represents the fission cross-section, $\nu_g$ is the average number of neutrons produced per fission event, $\Sigma^s_g$ represents the scatter cross-section, $\chi_g$ is the proportion of neutrons produced for each energy group per fission event and $N_g$ is the number of energy groups used. The subscript~$g$ denotes the particular energy group. The diffusion coefficient,~$D_g$, is defined as:
\begin{equation}\label{D_equation}
 D_g = \frac{1}{3(\Sigma^a_g + \Sigma^s_g)}\,.
\end{equation}
The eigenvalue, $\lambda$, is taken to be the reciprocal of $k_{\text{eff}}$ (i.e.~$\lambda = 1/k_{\text{eff}}$), where:
\begin{equation}
 k_{\text{eff}}=\frac{\text{number of neutrons in one generation}}{\text{number of neutrons in the preceding generation}} \ .
\end{equation}
Reflective and vacuum or bare surface boundary conditions can be implemented as follows:
\begin{align}\label{eq:diff-reflect}
D_g\left(\bm{n}\cdot \nabla \phi_g \right) &= 0 && \textit{(reflective)}\\
\label{eq:diff-bare}
-D_g\left( \bm{n}\cdot \nabla \phi_g  \right) & = \frac{1}{2} \phi_g && \textit{(vacuum or bare surface)}
 \end{align} 
where $\bm{n}$ is the outward-pointing normal to the~boundary. 

\subsection{Discretisation}
The diffusion equation in 2D can be discretised with finite volumes on a regular mesh of $N_x \times N_y$ cells as follows:
\begin{eqnarray}\label{eq:discretised_hfm}
& & -\frac{(D_{i-1,j,g}+D_{i,j,g})}{2\Delta x^2}\phi_{i-1,j,g} - 
    \frac{ (D_{i,j,g}+D_{i+1,j,g})}{2\Delta x^2}\phi_{i+1,j,g} - \frac{(D_{i,j-1,g}+D_{i,j,g})}{2\Delta y^2}\phi_{i,j-1,g}  \nonumber\\[2mm] 
& &-\frac{ (D_{i,j,g}+D_{i,j+1,g})}{2\Delta y^2}\phi_{i,j+1,g} + 
    \left(\frac{ (D_{i-1,j,g}+2D_{i,j,g}+D_{i+1,j,g})}{2\Delta x^2}+\frac{(D_{i,j-1,g}+2D_{i,j,g}+D_{i,j+1,g})}{2\Delta y^2}\right)\phi_{i,j,g}  \nonumber\\[2mm] 
 & & +
 {\Sigma^a}_{i,j,g}\phi_{i,j,g} \ + \ \sum_{\substack{g^{'}= 1\\g^{'}\neq g}}^{N_g}\Sigma^s_{i,j,g\rightarrow i,j,g^{'}}\phi_{i,j,g}
= \sum_{g^{'}=1}^{N_g}\Sigma^s_{i,j,g^{'}\rightarrow i,j,g}\phi_{i,j,g^{'}} +
\lambda\chi_g\sum_{g^{'}=1}^{N_g} \nu_{g^{'}} \Sigma^f_{i,j,g^{'}} \phi_{i,j,g^{'}}, \\[2mm]
& & \qquad \forall i\in\{2,3,\ldots,N_x-1\},\qquad \forall j\in\{2,3,\ldots,N_y-1\},\qquad \forall g\in\{1,2,\ldots,N_g\},\nonumber
\end{eqnarray}
where $\Delta x$ and $\Delta y$ are the uniform cell widths in the $x$ and $y$ directions respectively, $N_x$ and $N_y$ are the numbers of cells in the $x$ and $y$ directions respectively, the subscripts~$i$ and $j$ refer to the cells in the $x$ and $y$ directions respectively and $\phi_{i,j,g}$ represents the scalar flux of energy group $g$ in cell~$i,\,j$. This discretisation is equivalent to a finite difference discretisation. Boundary conditions are applied to the first and last cells in both the $x$ and $y$ directions, so Equation~\eqref{eq:discretised_hfm} is not solved for these cells. We want to apply the boundary conditions in such a way as to avoid changing the discretisation stencil near the boundaries to maximise the efficiency of the implementation. With this in mind, reflective boundary conditions for the left edge ($i=1$) can be enforced by the following constraints:
\begin{equation}\label{eq:BC_left}
    \phi_{1,j,g}=0 , \quad  D_{1,j,g} = - D_{2,j,g} \quad \forall j
\end{equation}
and for the right edge ($i=N_x$):
\begin{equation}\label{eq:BC_right}
    \phi_{N_x,j,g}=0 , \quad D_{N_x,j,g} = - D_{N_x-1,j,g}  \quad \forall j \,.
\end{equation}
Similar constraints can be applied to the top and bottom edges as required. 
This way of implementing the boundary conditions ensures that there is an average diffusivity of zero at the interface between boundary cells and their neightbours, and this avoids any diffusion occurring across the interface.  For bare surface boundary conditions (see Equation~\eqref{eq:diff-bare}), where the normal to the boundary is aligned with the $x$-direction, the absorption term is modified as follows:
\begin{equation}\label{eq:abs-bc-x}
\Sigma^a_{i,j,g}\leftarrow \Sigma^a_{i,j,g} + \frac{1}{2\Delta x}\,.
\end{equation}
For bare surface boundary conditions where the boundary is aligned with the $y$ direction: 
\begin{equation}\label{eq:abs-bc-y}
\Sigma^a_{i,j,g}\leftarrow \Sigma^a_{i,j,g} + \frac{1}{2\Delta y}\,.
\end{equation}
For cells that have both boundary conditions the following modification is made: 
\begin{equation}\label{eq:abs-bc-xy}
\Sigma^a_{i,j,g}\leftarrow \Sigma^a_{i,j,g} + \frac{1}{2\Delta x} + \frac{1}{2\Delta y}\,. 
\end{equation}
For the 5~point stencil associated with Equation~\eqref{eq:discretised_hfm}, the boundary conditions are implemented through one layer of `halo cells'. For higher order discretisations, with larger stencils, more layers of cells will be required to serve as halo cells.

Equation~\eqref{eq:discretised_hfm} and its associated boundary conditions are often written as:
\begin{equation}\label{disc_fiss}
 \bm{A}\bm{\phi}=\lambda \bm{B}\bm{\phi}.
\end{equation}
where the matrix~$\bm{A}$ contains the absorption, diffusion and scattering terms; matrix~$\bm{B}$ represents the fission terms; and the vector~$\bm{\phi}$ contains the values of the scalar flux for each cell in every energy group. In the following, we instead keep with the notation used thus far, which stores the unknown scalar flux of each energy group in a 2D array. 
Although this way of formulating the problem may be less familiar, the motivation will become clear in the following section, when we compare discretisation stencils to convolutional operators. Bearing this in mind, we rewrite the system in Equation~\eqref{eq:discretised_hfm} as
\begin{equation}\label{eq:discretised_hfm_rewritten}
\sum_{u=-l}^{l}\sum_{v=-l}^{l}a_{i,j,g}^{u,v} \, \phi_{i+u,j+v,g} =  s_{i,j,g}\,,\quad  \forall i\in\{2,3,\ldots,N_x-1\},\  \forall j\in\{2,3,\ldots,N_y-1\},\ \forall g\in\{1,2,\ldots,N_g\},
\end{equation}
where
\begin{eqnarray}
a_{i,j,g}^{u,v}&=&
\begin{cases}
\begin{array}{ll}
-\,\dfrac{(D_{i,j,g}+D_{i+u,j+v,g})}{2\Delta x^2} & \text{for}\ |u|=1,\,v=0\\
-\,\dfrac{(D_{i,j,g}+D_{i+u,j+v,g})}{2\Delta y^2} & \text{for}\ u=0,\,|v|=1\\
\dfrac{D_{i-1,j,g} +2D_{i,j,g}+D_{i+1,j,g}}{2\Delta x^2} + \dfrac{D_{i,j-1,g} +2D_{i,j,g}+D_{i,j+1,g}}{2\Delta y^2} + {\Sigma}^{as}_{i,j,g} & \text{for}\ u=0=v\\
0 & \text{for}\ |u|=1=|v| \\
\end{array}
\end{cases} \label{eq:define_aijuv} \\
{\Sigma}^{as}_{i,j,g} & =& {\Sigma}^a_{i,j,g} + \displaystyle{\sum_{\substack{g^{'}= 1\\g^{'}\neq g}}^{N_g}}\Sigma^s_{i,j,g\rightarrow i,j,g^{'}}\\
s_{i,j,g} & = & \sum_{g^{'}=1}^{g-1}\Sigma^s_{i,j,g^{'}\rightarrow i,j,g}\phi^{(k+1)}_{i,j,g^{'}}+\sum_{g^{'}=g}^{N_g}\Sigma^s_{i,j,g^{'}\rightarrow i,j,g}\phi^{(k)}_{i,j,g^{'}} + \lambda \chi_g\sum_{g^{'}=1}^{N_g} \nu_{g^{'}} \Sigma^f_{i,j,g^{'}} \phi^{(k)}_{i,j,g^{'}} \ . \label{eq:source_term}
\end{eqnarray}
As the stencil used in Equation~\eqref{eq:discretised_hfm} is a 5~point stencil (which can be written equivalently as a 3~by~3 stencil), the value of~$l$ in Equation~\eqref{eq:discretised_hfm_rewritten} is~1. The right-hand side of Equation~\eqref{eq:discretised_hfm_rewritten} can be determined by using a ``best guess" for $\phi_{i,j,g}$.  
This effectively linearises Equation~\eqref{eq:discretised_hfm_rewritten} which can now be solved by the Jacobi method:

\begin{equation}\label{eq:jacob_sol}
    \phi_{i,j,g}^{(k+1)} = \frac{1}{a_{i,j,g}^{0,0}}\left(s_{i,j,g}-\sum_{u=-l}^{l}\sum_{v=-l}^{l}a_{i,j,g}^{u,v} \, \phi_{i+u,j+v,g}^{(k)} +
    a_{i,j,g}^{0,0}\phi^{(k)}_{i,j,g}\right),
\end{equation}
%
where $2l+1$ is the width of the stencil, $k$ is the Jacobi iteration and $\{\{a_{i,j,g}^{u,v}\}_{u=-l}^{l}\}_{v=-l}^{l}$ represents the coefficients of the stencil used to calculate the scalar flux in cell $i,\,j$. The Jacobi method can be used for diagonally dominant systems, and given an initial guess, is solved for each diagonal component in turn. Iteration continues until the system converges~\citep{Acton}. The diagonal terms of the usual matrix-vector form of Equation~\eqref{eq:jacob_sol} (seen in Equation~\eqref{disc_fiss}) are now denoted by $a^{0,0}_{i,j,g}$ (for cell~$i,\,j$ and energy group~$g$), the remaining terms are the non-diagonal terms ($a^{u,v}_{i,j,g} \ \forall u,\,v$ such that $|u|=1$ and $v=0$, and $u=0$ and $|v|=1$), which are subtracted from the source term in Equation~\eqref{eq:jacob_sol}. 

\subsection{Implementing discretisations with convolutional neural networks}\label{sec:nn-alt}
\begin{figure}[!htb]
\centering
\scalebox{0.5}{\begin{tikzpicture}
\draw[step=1.0,black,thin] (0,0) grid (5,5);
\draw[blue,line width = 1mm] (1,1) rectangle ++ (3,3);
\draw[red,line width = 1mm] (2,2) rectangle ++ (1,1);
\foreach \x in {0.5,...,4.5}
        \node[font=\huge] at (0.5,\x) {1}; 
\foreach \x in {0.5,...,4.5}
        \node[font=\huge] at (1.5,\x) {2}; 
\foreach \x in {0.5,...,4.5}
        \node[font=\huge] at (2.5,\x) {5}; 
\foreach \x in {0.5,...,4.5}
        \node[font=\huge] at (3.5,\x) {4}; 
\foreach \x in {0.5,...,4.5}
        \node[font=\huge] at (4.5,\x) {1}; 

\draw[step=1.0,black,thin] (7,1) grid (10,4);
\node[] at (2.5,6) {Input ($N_x\ \times\ N_y$)};
\node[] at (8.5,6) {Filter};
\node[font=\huge] at (7.5,1.5) {0}; 
\node[font=\huge] at (7.5,2.5) {-1}; 
\node[font=\huge] at (7.5,3.5) {0}; 
\node[font=\huge] at (8.5,1.5) {-1}; 
\node[font=\huge] at (8.5,2.5) {4}; 
\node[font=\huge] at (8.5,3.5) {-1}; 
\node[font=\huge] at (9.5,1.5) {0}; 
\node[font=\huge] at (9.5,2.5) {-1}; 
\node[font=\huge] at (9.5,3.5) {0};

\node[font=\huge] at (6,2.5) {*};
\node[font=\huge] at (11,2.5) {=};
\draw[step=1.0,black,thin] (12,1) grid (15,4);
\node[] at (13.5,6) {Sum of Values};
\node[font=\huge] at (12.5,1.5) {0}; 
\node[font=\huge] at (12.5,2.5) {-2}; 
\node[font=\huge] at (12.5,3.5) {0}; 
\node[font=\huge] at (13.5,1.5) {-5}; 
\node[font=\huge] at (13.5,2.5) {20}; 
\node[font=\huge] at (13.5,3.5) {-5}; 
\node[font=\huge] at (14.5,1.5) {0}; 
\node[font=\huge] at (14.5,2.5) {-4}; 
\node[font=\huge] at (14.5,3.5) {0}; 
\draw[blue,line width = 1mm] (12,1) rectangle ++ (3,3);
\node[font=\huge] at (16,2.5) {=};
\draw[step=1.0,gray,thin] (17,0) grid (22,5);
\draw[red,line width = 1mm] (19,2) rectangle ++ (1,1);
\node[font=\huge] at (19.5,2.5) {4}; 
\node[] at (19.5,6) {Output};
\end{tikzpicture}}
\caption{A 3~by~3 convolutional filter which applies the discretised diffusion operator (a five-point finite volume stencil) in~2D to 9~cells. The filter is first applied to all the cells in the blue block on the left; the result of which can be seen in the blue block following the equals sign. The 9 values are then summed to give the value in the red block on the right which is the output value, which represents the value of the diffusion operator acting on the input approximated at the central cell.}
\label{fig:conv_filter}
\end{figure}
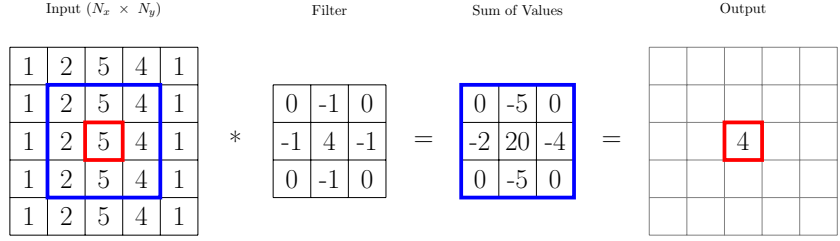

A convolutional layer of a neural network has a filter or kernel associated with it, which is a small grid (typically of dimension $3\times 3$, $5\times 5$ or $7\times 7$ for 2D filters) whose cells have values known as weights associated with them. The filter is applied to part of the input by multiplying the input value by the weight in the overlapping cells. The products are summed to produce the output. This is illustrated in Figure~\ref{fig:conv_filter}, where a filter acting on one part of the input data can be seen. 
This process of passing the filter over parts of the input data is repeated until the filter has passed over all the input data and all the output values are known. The action of a 2D convolutional layer on a 2D input can be written as follows 
\begin{equation}\label{eq:definition_convolutional_operation}
    x^{(k+1)}_{i,j} = \sum^l_{u=-l}\sum^l_{v=-l}w^{u,v}\,x^{(k)}_{i+u,j+v}, 
\end{equation}
where the input and output are 2D grids with components $x^{(k)}_{i,j}$ and $x^{(k+1)}_{i,j}$ respectively. The weights of the filter are represented by $w^{u,v}$ and the size of the filter is $(2l+1) \times (2l+1)$. The example in Figure~\ref{fig:conv_filter} corresponds to applying a discretised diffusion operator to an input for the case $\Delta x = 1 = \Delta y$ and a constant diffusivity ($\bm{D}_g$) of~1. For general grid sizes, using the notation in Equation~\eqref{eq:definition_convolutional_operation} and for a particular set of weights~$\bm{w}$, 
the discretised diffusion operator applied to the scalar flux of energy group~$g$ and cell~$i,j$ can be written as 
\begin{equation}\label{eq:Laplacian_standard}
    - \nabla^2\bm{\Phi}_g \Big|_{i,\,j}= \sum_{u=-l}^{l}\sum_{v=-l}^{l} w^{u,v}\phi_{i+u,j+v,g} = - \left( \frac{\phi_{i-1,j,g} - 2\phi_{i,j,g} 
+\phi_{i+1,j,g}}{\Delta x^2} 
    +\frac{\phi_{i,j-1,g} - 2\phi_{i,j,g} + \phi_{i,j+1,g}}{\Delta y^2} \right) \,,
\end{equation}
which is equivalent to 
\begin{equation}\label{eq:Laplacian_convolution}
   \sum_{u=-l}^{l}\sum_{v=-l}^{l} w^{u,v}\phi_{i+u,j+v,g} = 
 \sum_{\text{entries}}\begin{bmatrix}
 0
 & \frac{-1\phantom{-}}{\Delta y^2} & 0 \\
\frac{-1\phantom{-}}{\Delta x^2} & \frac{2}{\Delta x^2} +\frac{2}{\Delta y^2} &\frac{-1\phantom{-}}{\Delta x^2} \\
 0 & \frac{-1\phantom{-}}{\Delta y^2} & 0
 \end{bmatrix}\odot 
 \begin{bmatrix*}[l]
 \phi_{i-1,j+1,g} & \phi_{i,j+1,g} & \phi_{i+1,j+1,g} \\
 \phi_{i-1,j,g} & \phi_{i,j,g} & \phi_{i+1,j,g} \\
 \phi_{i-1,j-1,g} & \phi_{i,j-1,g} & \phi_{i+1,j-1,g} \\
 \end{bmatrix*} =: \bm{f}(\bm{\Phi}_g;\bm{w})\Big|_{i,j}. 
\end{equation}
where $l=1$, $\odot$ denotes the Hadamard product which performs entrywise multiplication, the symbol $\sum_{\text{entries}}$ denotes the summation of all the entries of a matrix (see Equation~\eqref{eq:definition_sum_entries}) and $\bm{f}$ represents the discrete convolution applied to the field $\bm{\Phi}_g$ by a $3 \times 3$ filter with weights~$\bm{w}$. The components inside the square brackets used in Equation~\eqref{eq:Laplacian_convolution} are ordered as if they are pixels in an image rather than components of a matrix.
Equation~\eqref{eq:Laplacian_standard} shows one way of writing a finite volume discretisation of the diffusion operator acting on a field~$\bm{\Phi}_g$ and  Equation~\eqref{eq:Laplacian_convolution} is exactly the same discretisation written as a convolution (using the Hadamard product). This illustrates how a discretisation scheme can be represented by a convolutional neural network.  

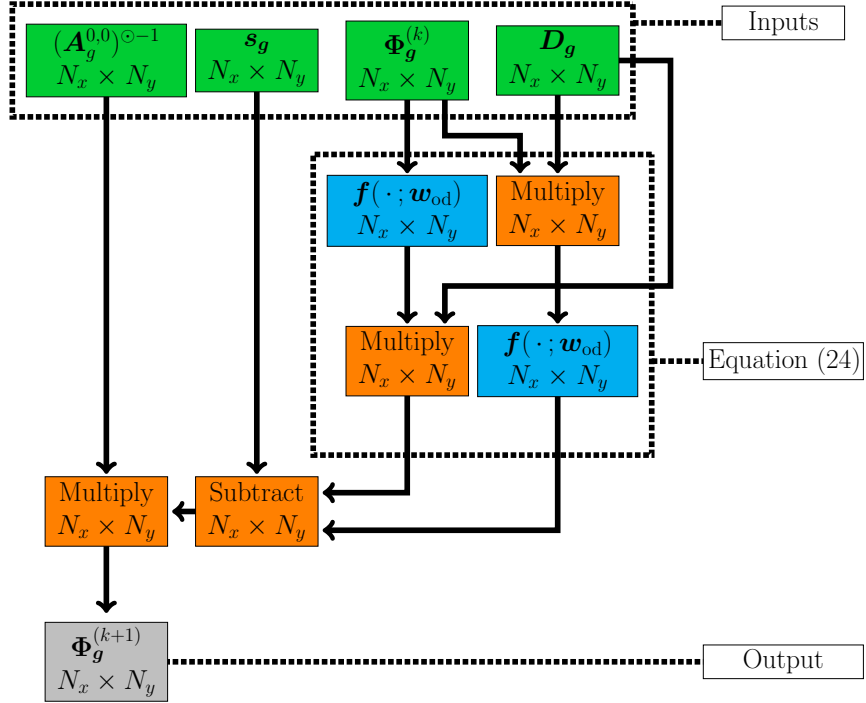
\begin{figure}[!htb]
\centering
\scalebox{0.5}{\begin{tikzpicture}



\draw [line width=1.5mm,-to] (12,12) -- (12,9);
\draw [line width=1.5mm,-to] (8,12) -- (8,9);
\draw [line width=1.5mm,-to] (9,12) -- (9,10) -- (11,10) -- (11,9);

\draw [line width=1.5mm,-to] (12,8) -- (12,5);
\draw [line width=1.5mm,-to] (8,8) -- (8,5);

\draw [line width=1.5mm,-to] (0,12) -- (0,1);
\draw [line width=1.5mm,-to] (4,0) -- (1.75,0);
\draw [line width=1.5mm,-to] (4,12) -- (4,1);

\draw[line width = 1.5mm, dotted] (14,13.5) -- (14,10.5) -- (-2.5,10.5) -- (-2.5,13.5) -- (14,13.5);
\draw[line width = 1.5mm, dotted] (14,13) -- (18,13);
\node[draw,font=\huge,text width = 3cm,align=center, fill = white] at (18,13) {Inputs};
\draw[line width = 1.5mm, dotted] (14.5,9.5) -- (5.5,9.5) -- (5.5,1.5) -- (14.5,1.5) -- (14.5,9.5);
\draw[line width = 1.5mm, dotted] (14.5,4) -- (18,4);
\node[draw,font=\huge,text width = 4cm,align=center, fill = white] at (18,4) {Equation~\eqref{eq:conv_off_diagonal}};
\draw[line width = 1.5mm, dotted] (0,-4) -- (18,-4);
\node[draw,font=\huge,text width = 4cm,align=center, fill = white] at (18,-4) {Output};

\draw [line width=1.5mm,-to] (12,4) -- (12,-0.5) -- (5.75,-0.5);
\draw [line width=1.5mm,-to] (8,4) -- (8,0.5) -- (5.75,0.5);
\draw [line width=1.5mm,-to] (12,12) -- (15,12) -- (15,6) -- (9,6) -- (9,5);

\draw [line width=1.5mm,-to] (0,0) -- (0,-2.7);

\node[draw,font=\huge,text width = 3cm,align=center,fill=green!80!blue] at (8,12) {$\bm{\Phi_g}^{(k)}$ \\ $N_{x}\times N_{y}$};

\node[draw,font=\huge,text width = 3cm,align=center,fill=green!80!blue] at (12,12) {$\bm{D_g}$ \\ $N_{x}\times N_{y}$};
\node[draw,font=\huge,text width = 3cm,align=center,fill=green!80!blue] at (4,12) {$\bm{s_g}$ \\ $N_{x}\times N_{y}$};
\node[draw,font=\huge,text width = 4cm,align=center,fill=green!80!blue] at (0,12) {$(\bm{A}_g^{0,0})^{\odot -1}$ \\ $N_{x}\times N_{y}$};

\node[draw,font=\huge,text width = 4cm,align=center,fill=cyan] at (8,8) {$\bm{f}(\,\cdot\,;\bm{w_{\text{od}}})$ \\ $N_{x}\times N_{y}$};

\node[draw,font=\huge,text width = 3cm,align=center,fill=orange] at (8,4) {Multiply \\ $N_{x}\times N_{y}$};

\node[draw,font=\huge,text width = 3cm,align=center,fill=orange] at (12,8) {Multiply \\ $N_{x}\times N_{y}$};

\node[draw,font=\huge,text width = 4cm,align=center,fill=cyan] at (12,4) {$\bm{f}(\,\cdot\,;\bm{w_{\text{od}}})$ \\ $N_{x}\times N_{y}$};

\node[draw,font=\huge,text width = 3cm,align=center,fill=orange] at (4,0) {Subtract \\ $N_{x}\times N_{y}$};

\node[draw,font=\huge,text width = 3cm,align=center,fill=orange] at (0,0) {Multiply \\ $N_{x}\times N_{y}$};

\node[draw,font=\huge,text width = 3cm,align=center,fill=lightgray] at (0,-4) {$\bm{\Phi_g}^{(k+1)}$  \\ $N_{x}\times N_{y}$};

\end{tikzpicture}}
\caption{Schematic of the neural network used for a single Jacobi iteration written as $\text{J}(\,\cdot\,)$ in  Equation~\eqref{eq:matrix_full}. This network performs a single Jacobi iteration on the flux of a single energy group. The inputs are the flux ($\bm{\Phi}_g^{(k)}$), representative source ($\bm{s}_g$), diffusion coefficients ($\bm{D}_g$) and the strictly diagonal coefficients ($\bm{A_g^{0,0}})^{\odot -1}$) (green boxes). A number of layer operations are performed, mathematical operations are shown in orange and convolutional operations are in cyan. The output is the flux of the next Jacobi iteration ($\bm{\Phi}_g^{(k+1)}$). Arrows originate from which layer the data originated and the end of an arrow indicates which layer takes that data as input. The dimensions of the layers are given on the second line of each box.}
\label{fig:jacobi_net}
\end{figure}

By comparing Equations~\eqref{eq:discretised_hfm_rewritten} and~\eqref{eq:definition_convolutional_operation}, we can see that the diffusion equation (with a spatially varying $\bm{D}_g$) cannot yet be written as a convolution layer, as the weights in Equation~\eqref{eq:discretised_hfm_rewritten} vary in space (for two cells $i,j$ and $i^*,j^*$, $a^{u,v}_{i,j} \neq a^{u,v}_{i^*,j^*}$), whereas in Equation~\eqref{eq:definition_convolutional_operation}, the weights do not depend on which part of the input data they are applied to (i.e.~$w^{u,v}$ is independent of $i,\,j$). By recalling that the diffusion operator in Equation~\eqref{eq:diff-eig} can be written as three terms all of which involve the Laplace operator (see \ref{sec:diffusion_operator_rewritten}), we can therefore write the diffusion operator as three convolutions:
\begin{align}
  - \nabla\cdot (D_g \nabla \phi_g) & = \frac{1}{2}\left( -\nabla^2(D_g \phi_g) - D_g  \nabla^2\phi_g + \phi_g\nabla^2D_g \right) && \textit{analytical form}\label{eq:analytical_form}\\[2mm]
  \bm{f}^{\text{Diff}}(\bm{\Phi}_g,\bm{D}_g; \bm{w}) &= \frac{1}{2}\left( \bm{f}(\bm{D}_g\odot\bm{\Phi}_g;\bm{w}) + \bm{D}_g\odot \bm{f}(\bm{\Phi}_g;\bm{w}) - \bm{\Phi}_g\odot \bm{f}(\bm{D}_g;\bm{w})\right) && \textit{discretised form}\label{eq:discretised_form} 
\end{align}
where $\bm{\Phi}_g$ is a $N_x \times N_y$ matrix containing all $\phi_{i,j,g}$ components, $\bm{D}_g$ is a $N_x \times N_y$ matrix containing all $D_{i,j,g}$ components and $\bm{f}$ represents the application of the convolutional layer with weights $\bm{w}$. Equation~\eqref{eq:discretised_form} serves as a definition of the diffusion convolution $\bm{f}^{\text{Diff}}$ and the weights~$\bm{w}$. The equivalence between this formulation of the finite volume discretisation of the diffusion operator and the standard formulation presented in Equation~\eqref{eq:discretised_hfm} can be seen in~\ref{sec:equivalence}.  The discretised diffusion equation can now be written for energy 
group~$g$ as: 
\begin{equation}\label{eq:conv_hfm}
\bm{f}^{\text{Diff}}(\bm{\Phi}_g, \bm{D}_g; \bm{w}) + \left({\bm{\Sigma}}^a_{g} \ + \ \sum_{\substack{g^{'}= 1\\g^{'}\neq g}}^{N_g}\bm{\Sigma}^s_{g\rightarrow g^{'}}\right)\odot\bm{\Phi}_{g} = \bm{s}_g, \quad \forall g\in\{1,2,\ldots,N_g\}, 
\end{equation} 
in which the source $\bm{s}_g$ for energy group $g$ also contains coupling terms between the energy groups other than~$g$. The terms $\bm{\Sigma}^a_{g}$ and $\bm{\Sigma}^s_{g\rightarrow g^{'}}$ represent matrices which contain the absorbtion and scatter cross-sections for each cell.   Equation~\eqref{eq:conv_hfm} can be solved with the Jacobi method as before. However, when implementing this, instead of using Equation~\eqref{eq:conv_hfm}, we rewrite this to use one fewer convolutional operation for efficiency.  
The term 
$\sum_{u=-l}^{l}\sum_{v=-l}^{l} a_{i,j,g}^{u,v}\,\phi_{i+u,j+v,g}^{(k)}-a_{i,j,g}^{0,0}\,\phi^{(k)}_{i,j,g} $ can be determined using a convolutional filter containing just the off-diagonal terms:
\begin{equation}\label{eq:conv_off_diagonal}
    \sum_{u=-l}^{l}\sum_{v=-l}^{l} a_{i,j,g}^{u,v}\,\phi_{i+u,j+v,g}^{(k)} -a_{i,j,g}^{0,0}\,\phi^{(k)}_{i,j,g}  \equiv \frac{1}{2}\left(\bm{D}_g\odot \bm{f}(\bm{\Phi}_{g}^{(k)};\bm{w_{\text{od}}})\left|_{{i,j,g}}\right.+ \bm{f}(\bm{D}_g\odot\bm{\Phi}_{g}^{(k)};\bm{w_{\text{od}}})\right)\Big|_{{i,j,g}}\ ,
\end{equation}
where $\bm{f}$ is a convolutional layer with weights $\bm{w_{\text{od}}}$:
\begin{gather}
\bm{w_{\text{od}}}
 = 
 \begin{bmatrix}
 0
 & \frac{-1\phantom{-}}{\Delta y^2} & 0 \\
\frac{-1\phantom{-}}{\Delta x^2} & 0 &\frac{-1\phantom{-}}{\Delta x^2} \\
 0 & \frac{-1\phantom{-}}{\Delta y^2} & 0
 \end{bmatrix}\,.
\end{gather}
with the central term of $\bm{w}$ set to zero in order to obtain this. The Jacobi method as written in Equation~\eqref{eq:jacob_sol} is therefore equivalent to:
\begin{equation}\label{eq:matrix_full}
    \bm{\Phi}_{g}^{(k+1)} = (\bm{A}_g^{0,0})^{\odot -1} \odot \left(\bm{s}_g -   \frac{1}{2}\left(\bm{D}_g\odot \bm{f}(\bm{\Phi}_{g}^{(k)};\bm{w_{\text{od}}})+ \bm{f}(\bm{D}_g\odot\bm{\Phi}_{g}^{(k)};\bm{w_{\text{od}}})\right)\right),
\end{equation} 
where $(\bm{A}_g^{0,0})^{\odot -1}$ is the Hadamard inverse~\citep{Reams1999} which is an $N_x \times N_y$ array whose $i^\text{th},j^\text{th}$ component is $\frac{1}{a^{0,0}_{i,j,g}}$ for energy group $g$. This equation can be written as a function~J, 
\begin{equation}
    \bm{\Phi}_g^{(k+1)} = \text{J}\left(\bm{\Phi}_g^{(k)},(\bm{A}_g^{0,0})^{\odot -1},\bm{s}_g,\bm{D}_g\right), 
\end{equation}
which calculates the updated solution after one Jacobi iteration. Figure~\ref{fig:jacobi_net} shows the architecture of this function, i.e., the neural network that solves one Jacobi iteration of the neutron transport problem as discretised in Equation~\eqref{eq:matrix_full}. Green boxes contain the inputs; blue boxes are convolutional layers; orange boxes are mathematical functions as layers; and the grey box is the output of the network. The second line in each box gives the dimension of the output of that box.

So far, we have described how to find the weights for the filters of convolutional layers that correspond to a finite volume discretisation of the neutron diffusion equation, solved with a Jacobi method. The approach described in this paper is not limited to the finite volume method, however, and for comparison, we also use a discretisation based on a new convolutional finite element method (ConvFEM)~\citep{Phillips2022-progress} of the diffusion operator. Using quadratic 9-noded rectangular elements, the $5 \times 5$ filter for this discretisation is given by
\begin{gather}\label{eq:quad_filt}
\bm{w}
 = 
 \frac{1}{900}
 \begin{bmatrix*}[r]
  -5&   50 &  -15 &   50&    -5 \\
   50&  -320   &   -660&      -320  &   50\\
  -15&  -660  &     3600 &      -660 &      -15\\
   50&  -320  &    -660  &    -320 & 50\\
  -5&   50&   -15&   50&  -5
 \end{bmatrix*} 
\end{gather}
\begin{equation}\label{eq:BC_dep}
    D_{\text{left}} = D_{\text{right}} = D_{\text{top}} =D_{\text{bottom}},
\end{equation}
and for the left side:
\begin{equation}\label{eq:BC_quad_dep}
    D_{\text{left}} = D_{i,j,g} \quad \forall i\in\{3,4\} \quad \forall j\in\{3,4,\ldots,N_y-2\},
\end{equation}
with corresponding constraints for the right, top and bottom sides. If Equation~\eqref{eq:BC_dep} holds and Equation~\eqref{eq:BC_quad_dep} holds for all sides then the boundary conditions for the left side, $i=1$, can be implemented with: \begin{equation}\label{eq:BC_left_quad}
    \phi_{1,j,g}=0 ,\quad \phi_{2,j,g}=0,  \quad  D_{\text{left}} = - D_{1,j,g} = - D_{2,j,g}, 
\end{equation}
and similar conditions for the other sides. Equation~\eqref{eq:BC_quad_dep} may not hold if two cells next to the boundary do not have the same value and thus this approach might not be used in this situation or one may use some sort of average.  
An alternative that works for all filter sizes is simply to set the values of the diffusion coefficient and the fluxes to be zero in the halo regions and then no addition to the absorption cross sections for the boundary condition are required. This effectively implements the $2\Delta x$ extrapolation boundary condition obtained using Equations~\eqref{eq:abs-bc-x}, \eqref{eq:abs-bc-y} and~\eqref{eq:abs-bc-xy}.

\subsection{Multigrid}\label{sec:multi-grid}
\begin{figure}[H]
    \centering
    \includegraphics[scale=0.5]{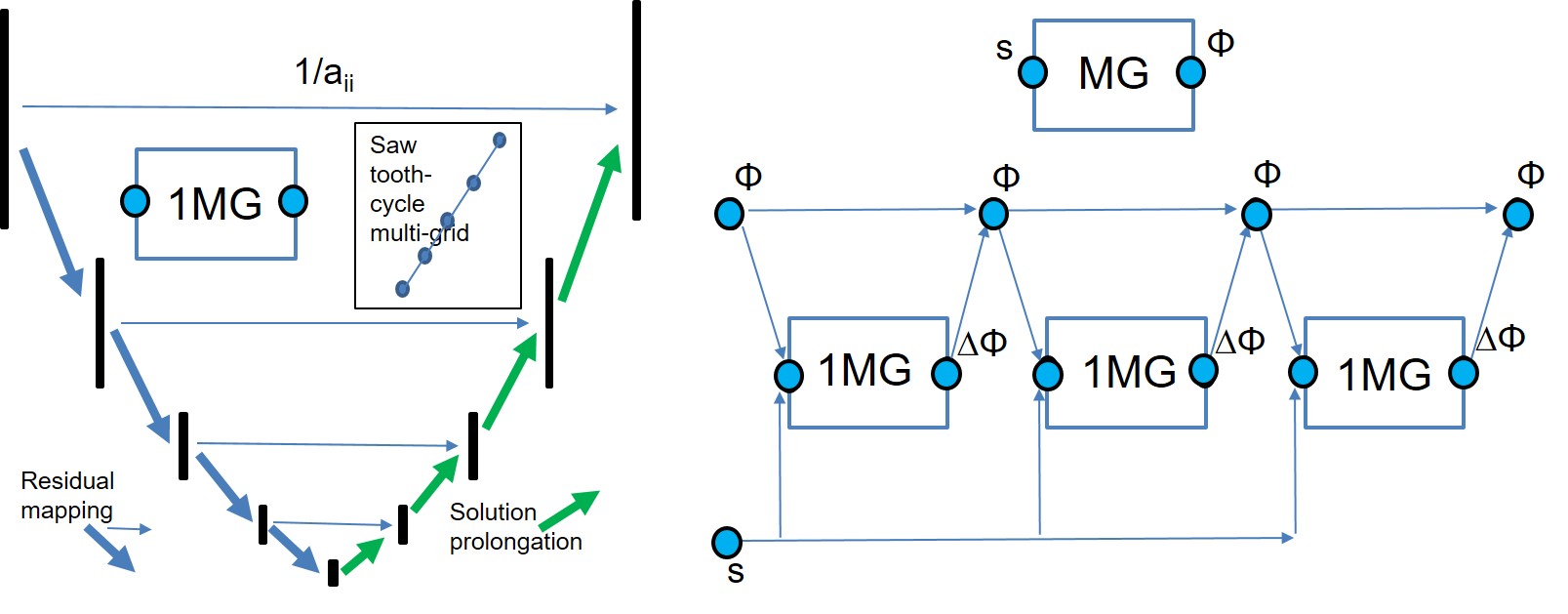}
    \caption{A schematic diagram showing the U-Net architecture (left) that is used to form a single multigrid sawtooth cycle. On the right, we can see how multiple cycles are brought together to form the overall solution method.}
    \label{fig:u-net-mg}
\end{figure}

In Figure~\ref{fig:u-net-mg} we show how the U-Net~\citep{Ronneberger2015} architecture has been repurposed to form a sawtooth multigrid method. 
\begin{figure}[!htb]
\centering
\scalebox{0.4}{\begin{tikzpicture}
\filldraw[cyan] (0,6) circle (8pt);
\filldraw[blue] (0,10) circle (8pt);
\node[font=\huge] at (7,11) {$\bm{r}^{(k)}_{\bm{1}} =\bm{s}_{\bm{1}} - \left(\bm{A}_{\bm{1}}^{0,0} \odot \bm{\Phi}_{\bm{1}}^{(k)} +  \frac{1}{2}\left( \bm{D}_{\bm{1}}\odot \bm{f}(\bm{\Phi}_{{\bm{1}}}^{(k)};\bm{w_{\text{od}}})+ \bm{f}(\bm{D}_{\bm{1}}\odot\bm{\Phi}_{{\bm{1}}}^{(k)};\bm{w_{\text{od}}})\right)\right)$};
\draw [line width=0.5mm,-to] (0,10) -- (0,8);
\node[font=\huge] at (0,7) {$\bm{r}^{(k)}_{\bm{1}}$};
\draw [line width=1.5mm,-to,cyan!60!black] (0.2,5.8) -- (3.6,4.2);
\node[font=\huge] at (6,5) {$\bm{r}^{(k)}_{\bm{2}} = \bm{f}^{\text{}}(\bm{r}^{(k)}_{\bm{1}};\bm{w_{R}})$};
\filldraw[cyan] (4,4) circle (8pt);
\draw [line width=1.5mm,-to,cyan!60!black] (4.2,3.8) -- (7.6,2.2);
\node[font=\huge] at (10,3) {$\bm{r}^{(k)}_{\bm{3}} = \bm{f}^{\text{}}(\bm{r}^{(k)}_{\bm{2}};\bm{w_{R}})$};
\filldraw[cyan] (8,2) circle (8pt);
\draw [line width=1.5mm,-to] (8.2,2) -- (13.6,2);

\filldraw[yellow] (14,2) circle (8pt);
\filldraw[teal] (18,4) circle (8pt);
\node[font=\huge] at (16,1) {$\Delta \bm{\Phi}^{(k)}_{\bm{3}} = \text{J}\left(\bm{0},(\bm{A}_{\bm{3}}^{0,0})^{\odot -1},\bm{r_3}^{(k)},\bm{D_2}\right)$};
\draw [line width=1.5mm,-to,teal!50!green] (14.2,2.2) -- (17.6,3.8);
\node[font=\huge] at (18,5) {$\widetilde{\Delta \bm{\Phi_2}}^{(k)}=\text{UpSamp}(\Delta \bm{\Phi}^{(k)}_{\bm{3}} )$};
\filldraw[yellow] (24,4) circle (8pt);
\draw [line width=1.5mm,-to] (18.2,4) --
(23.6,4);
\filldraw[teal] (28,6) circle (8pt);
\draw [line width=1.5mm,-to,teal!50!green] (24.2,4.2) -- (27.6,5.8);
\filldraw[yellow] (34,6) circle (8pt);
\draw [line width=1.5mm,-to] (28.2,6) -- (33.6,6);
\draw [line width=0.5mm,-to] (0.2,6) -- (27.6,6);
\draw [line width=0.5mm,-to] (4.2,4) -- (17.6,4);
\node[font=\huge] at (26,3) {$\Delta \bm{\Phi}^{(k)}_{\bm{2}} = \text{J}\left(\widetilde{\Delta \bm{\Phi_2}}^{(k)},(\bm{A}_{\bm{2}}^{0,0})^{\odot -1},\bm{r_2}^{(k)},\bm{D_2}\right)$};
\node[font=\huge] at (28,7) {$\widetilde{\Delta \bm{\Phi_1}}^{(k)}=\text{UpSamp}(\Delta \bm{\Phi}^{(k)}_{\bm{2}} )$};
\node[font=\huge] at (36,5) {$\Delta \bm{\Phi_1}^{(k)} = \text{J}\left(\widetilde{\Delta \bm{\Phi_1}}^{(k)} ,(\bm{A}_{\bm{1}}^{0,0})^{\odot -1},\bm{r_1}^{(k)},\bm{D_1}\right)$};
\filldraw[blue] (34,10) circle (8pt);
\node[font=\huge] at (34,11) {$\bm{\Phi_1}^{(k+1)} = \bm{\Phi}^{(k)}_{\bm{1}} + \Delta \bm{\Phi}^{(k)}_{\bm{1}}$};
\draw [line width=0.5mm,-to] (34,6.2) -- (34,9.6);
\draw [line width=0.5mm,-to] (34,10) -- (0.4,10);
\end{tikzpicture}}
\caption{Multigrid iteration, with the subscript indicating the resolution and superscript representing the multigrid iteration and $\text{J}(\,\cdot\,)$ representing a Jacobi iteration. The residual is calculated and restricted twice, indicated by the cyan nodes. These residuals are used with Jacobi smoothing, indicated by yellow nodes. After smoothing, prolongation is performed using UpSampling layers, indicated by teal nodes. After the finest level is reached, the flux is updated and the process repeats.}
\label{fig:multi_grid_it}
\end{figure}
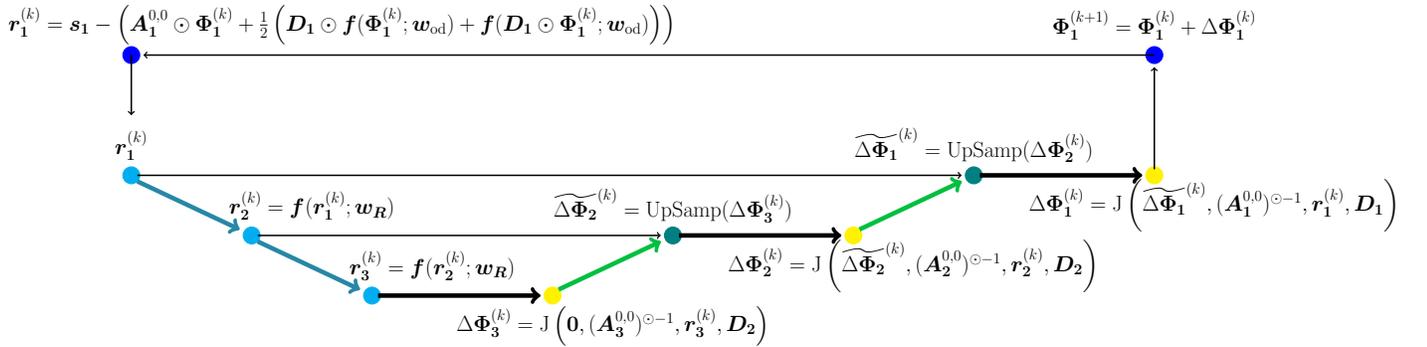
Figure~\ref{fig:multi_grid_it} shows a single multigrid iteration, using the U-Net, with two restrictions. Note that the subscript indicating energy group is no longer shown, the bold subscript now indicates the coarseness of the mesh, with~1 being the finest mesh. The residual ($\bm{r^k}$) is calculated using: 
\begin{equation}\label{eq:residual}
    \bm{r}^{(k)}_{\bm{1}} =\bm{s}_{\bm{1}} - \left(\bm{A}_{\bm{1}}^{0,0} \odot \bm{\Phi}_{{\bm{1}}}^{(k)} +  \frac{1}{2}\left( \bm{D}_{\bm{1}}\odot \bm{f}(\bm{\Phi}_{{\bm{1}}}^{(k)};\bm{w_{\text{od}}})+ \bm{f}(\bm{D}_{\bm{1}}\odot\bm{\Phi}_{{\bm{1}}}^{(k)};\bm{w_{\text{od}}})\right)\right)
\end{equation}
which is then 
restricted twice to ($\bm{r}^{(k)}_{\bm{2}}$) and ($\bm{r}^{(k)}_{\bm{3}}$). A Jacobi iteration is performed on the coarsest level (bold subscript~$3$) to determine $\Delta\bm{\Phi}^{(k)}_{\bm{3}}$, starting with an array of zeros. This is prolongated to estimate $\widetilde{\Delta \bm{\Phi_2}}^{(k)}$ which is then smoothed to  ${\Delta \bm{\Phi}^{(k)}_{\bm{2}}}$ with another Jacobi iteration using the residual of the next highest level. This repeats until the finest level is reached (bold subscript~$1$), where the flux is updated ($k+1$) and the process is repeated for a number of multigrid iterations. The restriction may be performed with the convolution:
\begin{equation}
    \bm{r}^{(k)}_{\bm{2}} = \bm{f}^{\text{}}(\bm{r}^{(k)}_{\bm{1}};\bm{w_{R}}),
\end{equation}
with filter weights:
\begin{gather}
\bm{w_{R}}
 =
 \begin{bmatrix}
 0.25 &  0.25  \\
 0.25  &  0.25 \\
 \end{bmatrix}.
\end{gather}
Upsampling layers can perform the role of prolongating the solution to a higher level. The upsampling operation simply copies the value from the coarser cell to the associated cells on the finer grid, which increases the dimensions of the data~\cite{chollet2015keras} and results in an approximation for the data on a finer mesh:
\begin{equation}
    \widetilde{\bm{\Phi_1}}^{(k)} = \text{UpSamp}( \bm{\Phi}_{\bm{2}}^{(k)})\,.  
\end{equation}

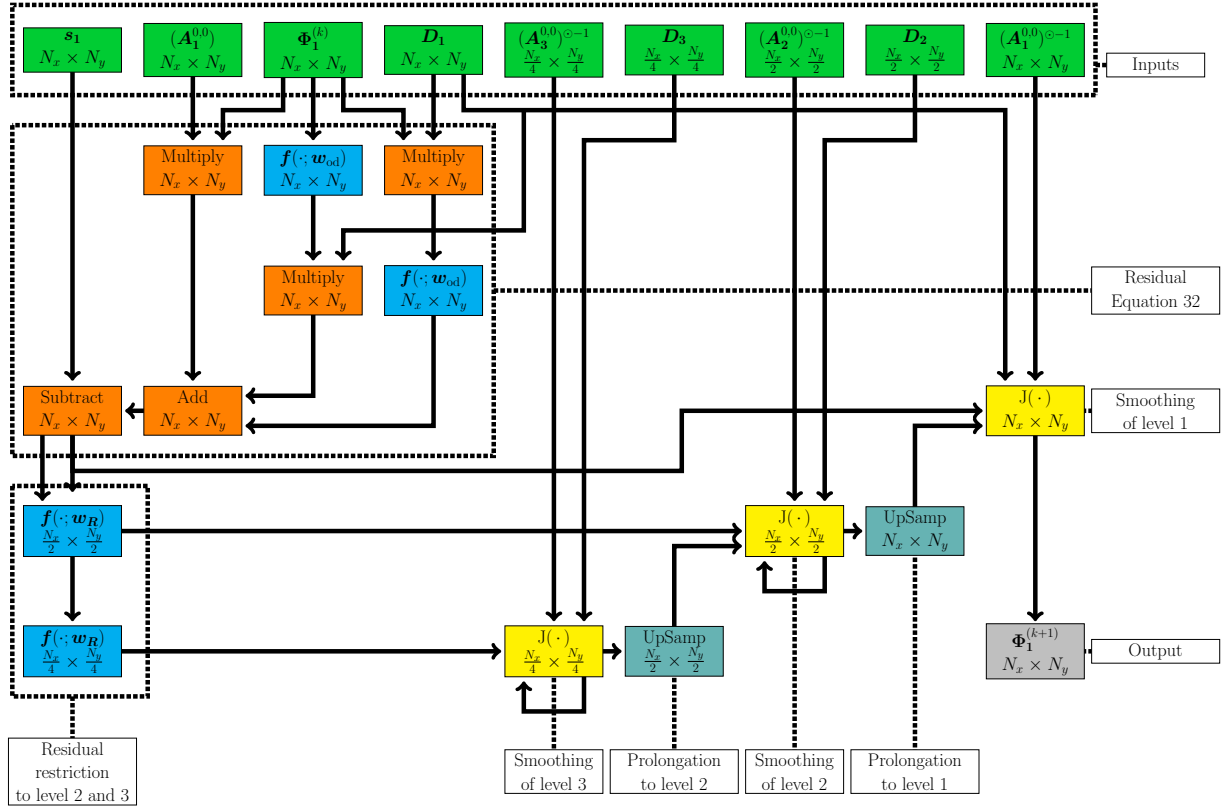
\begin{figure}[!htb]
\centering
\scalebox{0.4}{\begin{tikzpicture}
\draw [line width=1.5mm,-to] (12,12) -- (12,9);
\draw [line width=1.5mm,-to] (8,12) -- (8,9);
\draw [line width=1.5mm,-to] (9,12) -- (9,10) -- (11,10) -- (11,9);

\draw [line width=1.5mm,-to] (12,8) -- (12,5);
\draw [line width=1.5mm,-to] (8,8) -- (8,5);

\draw [line width=1.5mm,-to] (0,12) -- (0,1);
\draw [line width=1.5mm,-to] (4,0) -- (1.75,0);
\draw [line width=1.5mm,-to] (4,12) -- (4,9);

\draw [line width=1.5mm,-to] (4,8) -- (4,1);

\draw [line width=1.5mm,-to] (16,12) -- (16,-7);

\draw [line width=1.5mm,-to] (24,12) -- (24,-3);
\draw [line width=1.5mm,-to] (32,12) -- (32,1);

\draw [line width=1.5mm,-to] (17,-8) -- (17,-10) -- (15,-10) -- (15,-9);

\draw [line width=1.5mm,-to] (13,12) -- (13,10) -- (31,10) -- (31,1);

\draw [line width=1.5mm,-to] (28,12) -- (28,9) -- (25,9) -- (25,-3);

\draw [line width=1.5mm,-to] (20,12) -- (20,9) -- (17,9) -- (17,-7);

\draw [line width=1.5mm,-to] (16,-8) -- (18.25,-8);

\draw [line width=1.5mm,-to] (24,-4) -- (26.25,-4);

\draw [line width=1.5mm,-to] (25,-4) -- (25,-6) -- (23,-6) -- (23,-5);

\draw [line width=1.5mm,-to] (20,-8) -- (20,-4.5) -- (22.25,-4.5);

\draw [line width=1.5mm,-to] (28,-4) -- (28,-0.5) -- (30.25,-0.5);

\draw [line width=1.5mm,-to] (0,-8) -- (14.25,-8);

\draw [line width=1.5mm,-to] (-1,0) -- (-1,-3);
\draw [line width=1.5mm,-to] (0,-4) -- (0,-7);
\draw [line width=1.5mm,-to] (0,0) -- (0,-2) -- (20.25,-2) -- (20.25,0) -- (30.25,0);

\draw [line width=1.5mm,-to] (0,-4) -- (22.25,-4);

\draw [line width=1.5mm,-to] (32,0) -- (32,-7);

\draw [line width=1.5mm,-to] (7,12) -- (7,10) -- (5,10) -- (5,9);

\draw [line width=1.5mm,-to] (12,4) -- (12,-0.5) -- (5.75,-0.5);
\draw [line width=1.5mm,-to] (8,4) -- (8,0.5) -- (5.75,0.5);
\draw [line width=1.5mm,-to] (15,10) -- (15,6) -- (9,6) -- (9,5);

\draw [line width=1.5mm,-to] (0,0) -- (0,-2.7);

\draw[line width = 1.5mm, dotted] (34,10.5) -- (-2,10.5) -- (-2,13.5) -- (34,13.5) -- (34,10.5);
\draw[line width = 1.5mm, dotted] (34,11.5) -- (36,11.5);
\node[draw,font=\LARGE,text width = 3cm,align=center, fill = white] at(36,11.5) {Inputs};

\draw[line width = 1.5mm, dotted] (-2,9.5) -- (-2,-1.5) -- (14,-1.5) -- (14,9.5) --(-2,9.5);
\draw[line width = 1.5mm, dotted] (14,4) -- (36,4);
\node[draw,font=\LARGE,text width = 4cm,align=center, fill = white] at(36,4) {Residual \\ Equation~\ref{eq:residual}};

\draw[line width = 1.5mm, dotted] (-2,-2.5) -- (-2,-9.5) -- (2.5,-9.5) -- (2.5,-2.5) --(-2,-2.5);
\draw[line width = 1.5mm, dotted] (0,-9.5) -- (0,-12);
\node[draw,font=\LARGE,text width = 4cm,align=center, fill = white] at(0,-12) {Residual restriction \\ to level 2 and 3};

\draw[line width = 1.5mm, dotted] (16,-8) -- (16,-12);
\node[draw,font=\LARGE,text width = 3cm,align=center, fill = white] at(16,-12) {Smoothing of level 3};

\draw[line width = 1.5mm, dotted] (20,-8) -- (20,-12);
\node[draw,font=\LARGE,text width =4cm,align=center, fill = white] at(20,-12) {Prolongation to level 2};

\draw[line width = 1.5mm, dotted] (24,-4) -- (24,-12);
\node[draw,font=\LARGE,text width = 3cm,align=center, fill = white] at(24,-12) {Smoothing of level 2};

\draw[line width = 1.5mm, dotted] (28,-4) -- (28,-12);
\node[draw,font=\LARGE,text width = 4cm,align=center, fill = white] at(28,-12) {Prolongation to level 1};

\draw[line width = 1.5mm, dotted] (32,0) -- (36,0);
\node[draw,font=\LARGE,text width = 4cm,align=center, fill = white] at(36,0) {Smoothing of level 1};

\draw[line width = 1.5mm, dotted] (32,-8) -- (36,-8);
\node[draw,font=\LARGE,text width = 4cm,align=center, fill = white] at(36,-8) {Output};

\node[draw,font=\LARGE,text width = 3cm,align=center,fill=green!80!blue] at (8,12) {$\bm{\Phi}_{\bm{{1}}}^{(k)}$ \\ $N_{x}\times N_{y}$};

\node[draw,font=\LARGE,text width = 3cm,align=center,fill=green!80!blue] at (4,12) {$(\bm{A}_{\bm{1}}^{0,0})$ \\ $N_{x}\times N_{y}$};
\node[draw,font=\LARGE,text width = 3cm,align=center,fill=green!80!blue] at (0,12) {$\bm{s_1}$ \\ $N_{x}\times N_{y}$};
\node[draw,font=\LARGE,text width = 3cm,align=center,fill=green!80!blue] at (12,12) {$\bm{D_{1}}$ \\ $N_{x}\times N_{y}$};

\node[draw,font=\LARGE,text width = 3cm,align=center,fill=green!80!blue]
at (16,12) {$(\bm{A}_{\bm{3}}^{0,0})^{\odot -1}$ \\ $\frac{N_{x}}{4}\times \frac{N_{y}}{4}$};

\node[draw,font=\LARGE,text width = 3cm,align=center,fill=green!80!blue] at (20,12) {$\bm{D_{3}}$ \\ $\frac{N_{x}}{4}\times \frac{N_{y}}{4}$};

\node[draw,font=\LARGE,text width = 3cm,align=center,fill=green!80!blue]
at (24,12) {$(\bm{A}_{\bm{2}}^{0,0})^{\odot -1}$ \\ $\frac{N_{x}}{2}\times \frac{N_{y}}{2}$};

\node[draw,font=\LARGE,text width = 3cm,align=center,fill=green!80!blue] at (28,12) {$\bm{D_{2}}$ \\ $\frac{N_{x}}{2}\times \frac{N_{y}}{2}$};

\node[draw,font=\LARGE,text width = 3cm,align=center,fill=green!80!blue]
at (32,12) {$(\bm{A}_{\bm{1}}^{0,0})^{\odot -1}$ \\ $N_{x}\times N_{y}$};

\node[draw,font=\LARGE,text width = 3cm,align=center,fill=cyan] at (8,8) {$\bm{f}(\cdot;\bm{w_{\text{od}}})$ \\ $N_{x}\times N_{y}$};

\node[draw,font=\LARGE,text width = 3cm,align=center,fill=orange] at (8,4) {Multiply \\ $N_{x}\times N_{y}$};
\node[draw,font=\LARGE,text width = 3cm,align=center,fill=orange] at (4,8) {Multiply \\ $N_{x}\times N_{y}$};

\node[draw,font=\LARGE,text width = 3cm,align=center,fill=orange] at (12,8) {Multiply \\ $N_{x}\times N_{y}$};

\node[draw,font=\LARGE,text width = 3cm,align=center,fill=cyan] at (12,4) {$\bm{f}(\cdot;\bm{w_{\text{od}}})$ \\ $N_{x}\times N_{y}$};

\node[draw,font=\LARGE,text width = 3cm,align=center,fill=orange] at (4,0) {Add \\ $N_{x}\times N_{y}$};

\node[draw,font=\LARGE,text width = 3cm,align=center,fill=orange] at (0,0) {Subtract \\ $N_{x}\times N_{y}$};

\node[draw,font=\LARGE,text width = 3cm,align=center,fill=cyan] at (0,-4) {$\bm{f}(\cdot;\bm{w_{R}})$  \\ $\frac{N_{x}}{2}\times \frac{N_{y}}{2}$};

\node[draw,font=\LARGE,text width = 3cm,align=center,fill=cyan] at (0,-8) {$\bm{f}(\cdot;\bm{w_{R}})$  \\ $\frac{N_{x}}{4}\times \frac{N_{y}}{4}$};

\node[draw,font=\LARGE,text width = 3cm,align=center,fill=yellow] at (16,-8) {$\text{J}(\,\cdot\,)$  \\ $\frac{N_{x}}{4}\times \frac{N_{y}}{4}$};

\node[draw,font=\LARGE,text width = 3cm,align=center,fill=yellow] at (24,-4) {$\text{J}(\,\cdot\,)$  \\ $\frac{N_{x}}{2}\times \frac{N_{y}}{2}$};

\node[draw,font=\LARGE,text width = 3cm,align=center,fill=yellow] at (32,0) {$\text{J}(\,\cdot\,)$  \\ $N_{x}\times N_{y}$};

\node[draw,font=\LARGE,text width = 3cm,align=center,fill=teal!60!] at (28,-4) {UpSamp  \\ $N_{x}\times N_{y}$};

\node[draw,font=\LARGE,text width = 3cm,align=center,fill=teal!60!] at (20,-8) {UpSamp  \\ $\frac{N_{x}}{2}\times \frac{N_{y}}{2}$};

\node[draw,font=\LARGE,text width = 3cm,align=center,fill=lightgray] at (32,-8) {$\bm{\Phi_{1}}^{(k+1)}$  \\ $N_{x}\times N_{y}$};

\end{tikzpicture}}
\caption{Multigrid network, MG$(\,\cdot\,)$, representing a single multigrid iteration. This network performs a single multigrid iteration on the flux of a single energy group. Takes the flux ($\bm{\Phi}_{\bm{1}}^{(k)}$), representative source ($\bm{s_1}$), diffusion coefficients ($\bm{D_1}$) and the strictly diagonal coefficients ($\bm{A}_{\bm{1}}^{0,0}$), along with the coarser resolution coefficients, as inputs (green boxes). A number of layer operations are performed, mathematical operations in orange, convolutional passes in cyan, sub-model operations in yellow and upsampling in teal. The sub-models can be iterated on multiple times. Finally it outputs the flux of the next multigrid iteration flux ($\bm{\Phi}_{\bm{1}}^{(k+1)}$).  Arrow origins show which layer the data originated and the end of the arrow shows which layer takes that data as input. Dimensions of layers are given on the second line of each box.}
\label{fig:multi_grid_net}
\end{figure}

Figure~\ref{fig:multi_grid_net} shows how the multigrid method can be represented by a neural network. Green boxes contain the inputs, blue boxes are convolutional layers, orange boxes are mathematical functions as layers, yellow boxes are sub-networks, teal boxes are upsampling layers and the grey box is the output of the network. The second line in each box is the dimension of the output. 
This can be written as:
\begin{equation}\label{eq:multigroup_net_eq}
    \bm{\Phi}_{\bm{1}}^{(k+1)} = \text{MG}\left(\bm{\Phi}_{\bm{1}}^{(k)},\bm{s}_{\bm{1}},(\bm{A}_{\bm{1}}^{0,0})^{},(\bm{A}_{\bm{1}}^{0,0})^{\odot -1}, (\bm{A}_{\bm{2}}^{0,0})^{\odot -1},(\bm{A}_{\bm{3}}^{0,0})^{\odot -1},\bm{D_1},\bm{D_2},\bm{D_3}\right), 
\end{equation}
and for a single energy group~$g$: 
\begin{equation}\label{eq:multigroup_net_eq_1g}
    \bm{\Phi}_{\bm{1}g}^{(k+1)} = \text{MG}_g\left(\bm{\Phi}_{\bm{1}g}^{(k)},\bm{s}_{\bm{1}g},(\bm{A}_{\bm{1}g}^{0,0})^{},(\bm{A}_{\bm{1}g}^{0,0})^{\odot -1}, (\bm{A}_{\bm{2g}}^{0,0})^{\odot -1},(\bm{A}_{\bm{3}g}^{0,0})^{\odot -1},\bm{D}_{\bm{1}g},\bm{D}_{\bm{2}g},\bm{D}_{\bm{3}g}\right), 
\end{equation}
where $\text{MG}(\,\cdot\,)$ is a function that calculates the result of one sawtooth multigrid iteration 
and many of these iterations are strung together to form the final solution, see Figure~\ref{fig:u-net-mg}.  $\text{MG}_g(\,\cdot\,)$ is the multigrid iteration applied to energy group $g$, indicated by the subscript.

It should be noted that the diffusion coefficients and other material properties are mapped to a coarser grid using a harmonic average before the discretisations are formed on the coarser grids.  The same discretisation is used at each multigrid level but with different cell sizes.

\subsection{Multi-group network}
The multigrid function and network, given by Equation~\eqref{eq:multigroup_net_eq} and Figure~\ref{fig:multi_grid_net} respectively, show how a single multigrid iteration may be applied to a single energy group $g$. Multi-group problems must have balanced scattering terms, achieved through iterating until the terms balance. Equation~\eqref{eq:source_term} shows how a block Gauss-Seidel approach  is used when constructing $s_{i,j,g}$. The scattering term, $\Sigma^s$, is constructed using the most recent flux information, achieved by resolving each energy group sequentially. 

\begin{figure}[!htb]
\centering
\scalebox{0.4}{\begin{tikzpicture}

\draw [line width=1.5mm,] (0,12) -- (0,10) -- (6,10);

\draw [line width=1.5mm,] (6,12) -- (6,10);

\draw [line width=1.5mm,-to] (5,10) -- (5,8.5) -- (3,8.5);

\draw [line width=1.5mm,-to] (12,12) -- (12,7.5) -- (3,7.5);

\draw [line width=1.5mm,-to] (12,12) -- (12,4.5) -- (3,4.5);

\draw [line width=1.5mm,-to] (-1,0.5) -- (-4,0.5) -- (-4,4)-- (-3,4);

\draw [line width=1.5mm,-to] (5,-7.5) -- (2,-7.5) -- (2,-4)-- (3,-4);

\draw [line width=1.5mm,-to] (23,-15.5) -- (20,-15.5) -- (20,-12)-- (21,-12);


\draw [line width=1.5mm,-to] (0,0) -- (0,-14.5);
\draw [line width=1.5mm,-to] (24,-16) -- (3,-16);

\draw [line width=1.5mm] (15,-10) -- (15,-16);
\draw [line width=1.5mm] (6,-8) -- (6,-16);

\draw [line width=1.5mm] (30,12) -- (30,10) -- (18,10);
\draw [line width=1.5mm] (30,8) -- (30,6) -- (18,6);
\draw [line width=1.5mm] (30,4) -- (30,2) -- (18,2);
\draw [line width=1.5mm] (24,4) -- (24,2);
\draw [line width=1.5mm] (24,8) -- (24,6);
\draw [line width=1.5mm] (24,12) -- (24,10);
\draw [line width=1.5mm] (24,12) -- (24,10);

\draw [line width=1.5mm] (18,12) -- (18,2);
\draw [line width=1.5mm,-to] (18,3.5) -- (3,3.5);

\draw [line width=1.5mm,-to] (0,8) -- (0,5);]
\draw [line width=1.5mm,-to] (0,4) -- (0,1);
\draw [line width=1.5mm,-to] (0,0) -- (3.25,0);

\draw [line width=1.5mm,-to] (6,0) -- (6,-3);]
\draw [line width=1.5mm,-to] (6,-4) -- (6,-7);
\draw [line width=1.5mm,-to] (6,-8) -- (11.25,-8);

\draw [line width=1.5mm,-to] (24,-12) -- (24,-15);]
\draw [line width=1.5mm,-to] (24,-8) -- (24,-11);]

\draw [line width=1.5mm,-to] (19,-8) -- (21.25,-8);
\draw[line width = 1.5mm, dotted] (-3,13.5) -- (-3,10.5) -- (21,10.5) -- (21,1) --(33,1) -- (33,13.5) -- (-3,13.5);
\draw[line width = 1.5mm, dotted] (-3,12)-- (-7,12);
\node[draw,font=\huge,text width = 4cm,align=center, fill = white] at(-7,12) {Inputs};

\draw[line width = 1.5mm, dotted] (-2.75,9.5) -- (-2.75,-2) -- (2.75,-2) -- (2.75,9.5)-- (-2.75,9.5);
\draw[line width = 1.5mm, dotted] (-2.75,5)-- (-7,5);
\node[draw,font=\huge,text width = 4cm,align=center, fill = white] at(-7,5) {Resolve Energy Group 1};
\draw[line width = 1.5mm, dotted] (3.25,1.5) -- (3.25,-10) -- (8.75,-10) -- (8.75,1.5)-- (3.25,1.5);
\draw[line width = 1.5mm, dotted] (3.25,-5)-- (-7,-5);
\node[draw,font=\huge,text width = 4cm,align=center, fill = white] at(-7,-5) {Resolve Energy Group 2};

\draw[line width = 1.5mm, dotted] (10,-10) --(20,-10) -- (20,1)--  (10,1)-- (10,-10);
\node[draw,font=\huge,text width = 7cm,align=center, fill = white] at(15,-2) {Resolve Energy Groups $3, 4, \dots, (N_{g}-1)$};

\draw[line width = 1.5mm, dotted] (21.25,-6.5) -- (21.25,-18) -- (26.75,-18) -- (26.75,-6.5)-- (21.25,-6.5);
\draw[line width = 1.5mm, dotted] (24,-6.5)-- (24,-4.5);
\node[draw,font=\huge,text width = 4cm,align=center, fill = white] at(24,-4.5) {Resolve Energy Group $N_{g}$};

\draw[line width = 1.5mm, dotted] (0,-16)-- (-7,-16);
\node[draw,font=\huge,text width = 4cm,align=center, fill = white] at(-7,-16) {Output: \\ Concatenated energy groups};

\node[draw,font=\huge,text width = 5cm,align=center,fill=green!80!blue] at (0,12) {$\bm{\Sigma^s}$ \\ $N_{x}\times N_{y}\times N_{g}\times N_{g}$};

\node[draw,font=\huge,text width = 5cm,align=center,fill=green!80!blue] at (12,12) {$\bm{\Phi}_{}^{(k)}$ \\ $N_{x}\times N_{y}\times N_{g}$};

\node[draw,font=\huge,text width = 5cm,align=center,fill=green!80!blue] at (18,12) {$\bm{A}_{\bm{2}}^{0,0}$ \\ $N_{x}\times N_{y}\times N_{g}$};
\node[draw,font=\huge,text width = 5cm,align=center,fill=green!80!blue] at (6,12) {$\bm{s}_{\text{fiss}}$ \\ $N_{x}\times N_{y}\times N_{g}$};
\node[draw,font=\huge,text width = 5cm,align=center,fill=green!80!blue] at (24,12) {$\bm{D_{1}}$ \\ $N_{x}\times N_{y} \times N_{g}$};

\node[draw,font=\huge,text width = 5cm,align=center,fill=green!80!blue]
at (30,12) {$(\bm{A}_{\bm{1}}^{0,0})^{\odot -1}$ \\ $\frac{N_{x}}{4}\times \frac{N_{y}}{4}\times N_{g}$};

\node[draw,font=\huge,text width = 5cm,align=center,fill=green!80!blue] at (24,4) {$\bm{D_{3}}$ \\ $\frac{N_{x}}{4}\times \frac{N_{y}}{4}\times N_{g}$};

\node[draw,font=\huge,text width = 5cm,align=center,fill=green!80!blue]
at (30,8) {$(\bm{A}_{\bm{2}}^{0,0})^{\odot -1}$ \\ $\frac{N_{x}}{2}\times \frac{N_{y}}{2}\times N_{g}$};

\node[draw,font=\huge,text width = 5cm,align=center,fill=green!80!blue] at (24,8) {$\bm{D_{2}}$ \\ $\frac{N_{x}}{2}\times \frac{N_{y}}{2}\times N_{g}$};

\node[draw,font=\huge,text width = 5cm,align=center,fill=green!80!blue]
at (30,4) {$(\bm{A}_{\bm{3}}^{0,0})^{\odot -1}$ \\ $N_{x}\times N_{y}\times N_{g}$};

\node[draw,font=\huge,text width = 5cm,align=center,fill=yellow]
at (0,8) {Create $s_{g=1}$\\ $N_{x}\times N_{y}$};

\node[draw,font=\huge,text width = 5cm,align=center,fill=yellow]
at (0,4) {$\text{MG}_{g=1}(\,\cdot\,)$\\ $N_{x}\times N_{y}$};

\node[draw,font=\huge,text width = 5cm,align=center,fill=lightgray] at (0,0) {$\bm{\Phi}_{g=1}^{(k+1)}$  \\ $N_{x}\times N_{y}$};

\node[draw,font=\huge,text width = 5cm,align=center,fill=yellow]
at (6,0) {Create $s_{g=2}$\\ $N_{x}\times N_{y}$};

\node[draw,font=\huge,text width = 5cm,align=center,fill=yellow]
at (6,-4) {$\text{MG}_{g=2}(\,\cdot\,)$\\ $N_{x}\times N_{y}$};

\node[draw,font=\huge,text width = 5cm,align=center,fill=lightgray] at (6,-8) {$\bm{\Phi}_{\bm{}g=2}^{(k+1)}$  \\ $N_{x}\times N_{y}$};

\node[draw,font=\huge,text width = 5cm,align=center,fill=yellow]
at (24,-8) {Create $s_{g=N_{g}}$\\ $N_{x}\times N_{y}$};

\node[draw,font=\huge,text width = 5cm,align=center,fill=yellow]
at (24,-12) {$\text{MG}_{g=N_{g}}(\,\cdot\,)$\\ $N_{x}\times N_{y}$};

\node[draw,font=\huge,text width = 5cm,align=center,fill=lightgray] at (24,-16) {$\bm{\Phi}_{\bm{}g=N_{g}}^{(k+1)}$  \\ $N_{x}\times N_{y}$};

\node[draw,font=\huge,text width = 5cm,align=center,fill=lightgray] at (0,-16) {$\bm{\Phi_{}}^{(k+1)}$  \\ $N_{x}\times N_{y}\times N_{g}$};

\filldraw[black] (12,-8) circle (16pt);
\filldraw[black] (15,-8) circle (16pt);
\filldraw[black] (18,-8) circle (16pt);
\end{tikzpicture}}
\caption{Multi-group network representing a single multi-group iteration. This network performs a single multi-group iteration on the flux of all energy groups. Takes the flux ($\bm{\Phi_{1}}^{(k)}$), scattering cross-sections $\Sigma_s$, source fission term ($\bm{s}_{\text{fiss}}$), diffusion coefficients ($\bm{D_1}$) and the strictly diagonal coefficients ($\bm{A}_{\bm{1}}^{0,0}$), along with the coarser resolution coefficients, as inputs (green boxes). Each energy group is updated sequentially, first through updating the source term for a specific energy group and then performing a number of multigrid sub-model iterations. The updated flux for an energy group is then passed onto subsequent energy groups. Left out of the figure for clarity, the inputs (green boxes) are all connected to subsequent sub-models (yellow boxes). Finally, it outputs the flux of the next multi-group iteration flux ($\bm{\Phi_{1}}^{(k+1)}$).  Arrow origins show which layer the data originated and the end of the arrow shows which layer takes that data as input. Dimensions of layers are given on the second line of each box.}
\label{fig:multi_group_net}
\end{figure}
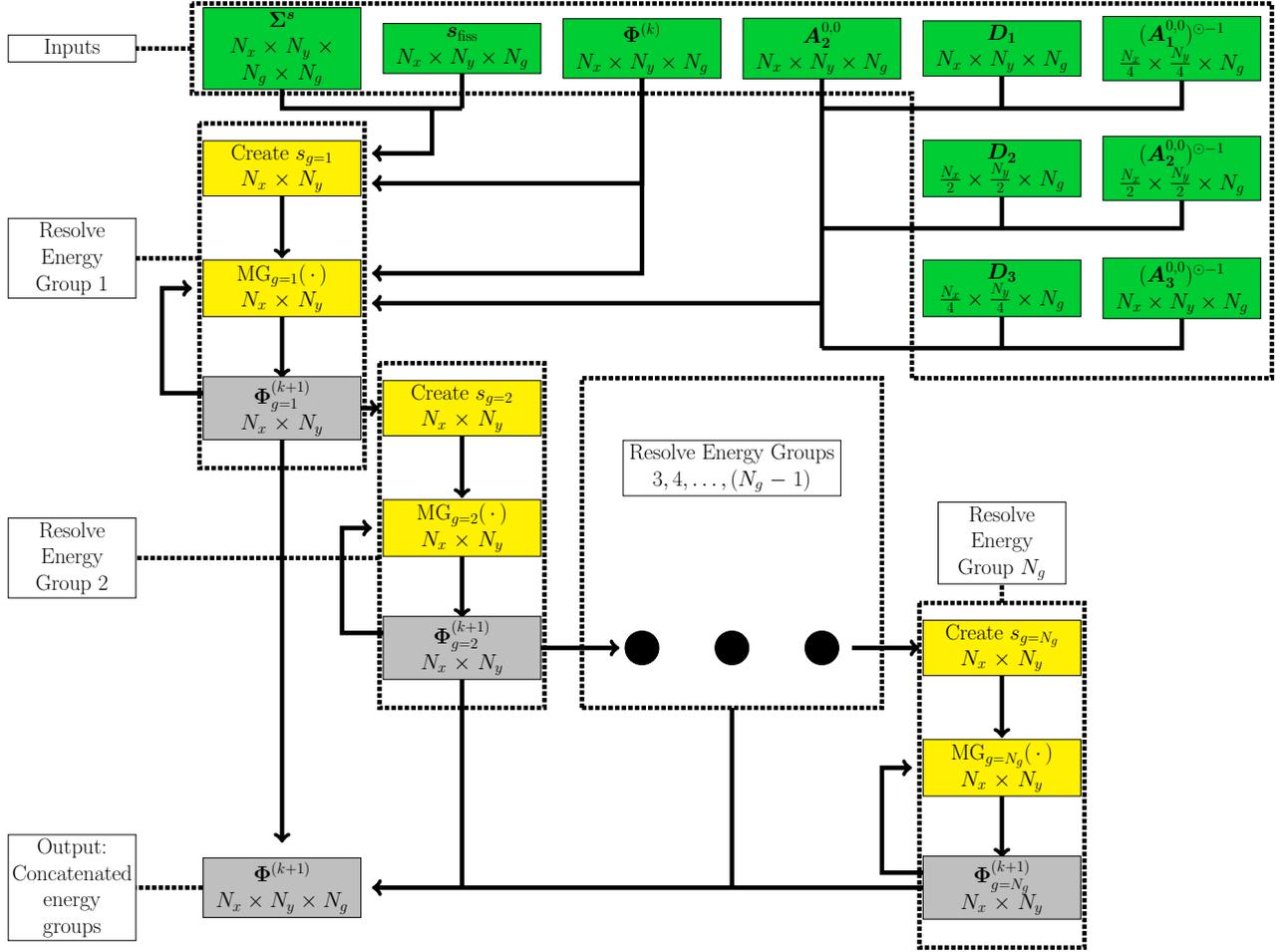
Figure~\ref{fig:multi_group_net} shows how energy groups can be resolved using a block Gauss-Seidel approach within a neural network. Green boxes represent inputs, yellow boxes represent sub-networks and grey boxes represent outputs. For clarity, the green outputs are only shown as being linked to the first energy group but would be linked to all subsequent energy groups. $\bm{s}_g$ is a vector containing  $\Sigma^s$, $\bm{s_\text{fiss}}$ and $\bm{\Phi}^{(k)}$ and is formed using Equation~\eqref{eq:source_term}. $\bm{s}_g$ is then used in the MG sub-model to resolve for $\bm{\Phi}_g$, repeating for a number of multigrid iterations until:
\begin{equation}
    \bm{\Phi}_g^{(k+1)} \approx \bm{\Phi}_g^{(k)} .
\end{equation}

$\bm{\Phi}_g$ is then used in  $\bm{s}_{g'}$ where $g' > g$. Once all energy groups have been resolved they can be concatenated to form $\bm{\Phi}^{(k+1)}$. This is repeated until:
\begin{equation}
    \bm{\Phi}^{(k+1)} \approx \bm{\Phi}^{(k)} .
\end{equation}

An alternative to the Gauss-Seidel approach would be to use the Jacobi approach to resolve all energy groups simultaneously, which could be achieved by using the multigrid network alone, as described in Section~\ref{sec:multi-grid}. This is performed by passing all $N_g$ energy groups to the MG network at the same time, only updating $\bm{s}$ outside of this. The source term in Equation~\eqref{eq:source_term} instead changes to:

\begin{equation}
    s_{i,j,g}  = \sum_{g^{'}=1}^{N_g}\Sigma^s_{i,j,g^{'}\rightarrow i,j,g}\phi^{(k)}_{i,j,g^{'}}  + \lambda \chi_g\sum_{g^{'}=1}^{N_g} \nu_{g^{'}} \Sigma^f_{i,j,g^{'}} \phi^{(k)}_{i,j,g^{'}}\,.
\end{equation}

Equation~\eqref{disc_fiss} is an eigenvalue problem so $\lambda$ needs to be determined. An approximation is used (usually $\lambda=1$) and the fission term is passed to the multi-group network where
\begin{equation}
    \bm{s}_{\text{fiss},g} = \lambda \chi_g\sum_{g^{'}=1}^{N_g} \nu_{g^{'}} \bm{\Sigma^f}_{g^{'}} \bm{\Phi}^{(k)}_{g^{'}},
\end{equation}
for each energy group~$g$ and $\bm{s}_{\text{fiss}}$ is an array containing all $g$ of $\bm{s}_{\text{fiss},g}$. The power method~\cite{Golub1996} is the method chosen here to determine the dominant eigenvalue for this problem. The implementation of the power method used here is the same as~\cite{phillips2021} and operates outside of the multi-group network.

\section{Results}\label{results}
The approach described in this paper is demonstrated on two test cases: a fuel assembly and a reactor core, both based on the KAIST benchmark~\citep{kaist2000}. For the fuel assembly, two configurations are investigated (control rods fully withdrawn and fully inserted). Results for a finite volume discretisation of the 2D neutron diffusion equation are generated by a neural network with pre-determined weights and compared with a results from a traditional Fortran implementation. A neural network solution of a discretisation based on the quadratic finite element method, ConvFEM~\cite{Phillips2022-progress}, is also presented. For the reactor core, the cross-sections are taken from the KAIST benchmark, and a grid of $3 \times 3$ fuel assemblies are used to make up one quarter of the core. Results are presented from a finite volume discretisation of the neutron diffusion equation using a neural network. All the neural networks in this section were implemented in python using Keras~\cite{chollet2015keras} with the TensorFlow backend~\cite{tensorflow2015}. 

\subsection{Fuel Assembly - Geometry and Configuration}\label{sec:fuel_assem_geom}
The geometry of the UOX fuel assembly based on the KAIST benchmark~\citep{kaist2000} can be seen in Figure~\ref{fig:fuel_assembly_mesh}. It consists of a $17 \times 17$ lattice containing 264~UOX fuels rods with guide tubes in the remaining 25~lattice-cells which can be filled with either moderator or control rods. We consider two configurations of the assembly. In the first configuration, all 25~of these lattice-cells are filled with moderator, representing a system where the control rods are fully withdrawn. In the second configuration, all 25~of the remaining lattice-cells are control rods, representing a system where the control rods are fully inserted. 
 Two computational grids are used, with either $20 \times 20$ cells or $10 \times 10$ cells within each lattice-cell. The higher resolution grid is used for the fuel assembly test case and the coarser grid is used when modelling the whole reactor (see Section~\ref{sec:reactor_core_results}). For the $20 \times 20$ case, there is a total of $115,600$ computational cells in the lattice with $1,364$ of these forming the boundaries (i.e.~as halo cells or ghost cells). The energy is discretised into seven groups, meaning that the fuel assembly has $818,720$ degrees of freedom. Each side of the fuel assembly is of length \SI{21.42}{cm} meaning each computational cell measures \SI{0.063}{cm} $\times$ \SI{0.063}{cm}. Each side of the fuel assembly has vacuum boundary conditions applied to it. 
 
\begin{figure}[H]
\centering
\scalebox{0.5}{\begin{tikzpicture}
\draw[step=1.0,black,thin] (0,0) grid (17,17);
\foreach \x in {0.5,3.5,6.5}
\foreach \y in {0.5,3.5,...,12.5}
\draw[black] (2+\y,5+\x) circle (7pt); 
\foreach \x in {0.5,3.5,6.5}
\foreach \y in {0.5,3.5,...,12.5}
\draw[black] (5+\x,2+\y) circle (7pt); 
\foreach \x in {3.5,13.5}
\foreach \y in {3.5,13.5}
\draw[black] (\x,\y) circle (7pt); 

\draw[step=1.0,black,thin] (18,11) grid (19,12);
\node[font=\huge,text width = 3cm] at (21,11.5) {Fuel Rod};
\draw[step=1.0,black,thin] (18,5) grid (19,6);
\draw[black] (18.5,5.5) circle (7pt); 
\node[font=\huge,text width = 5cm] at (22,5.5) {Control Rod or\\ Moderator};
\node[font=\huge,text width = 5cm] at (-5,5.5) {\phantom{Control Rod or}};


\end{tikzpicture}}
\caption{Geometry of UOX fuel assembly with laatice-cells containing fuel rods or guide tubes with either moderator or control rods.}
\label{fig:fuel_assembly_mesh}
\end{figure}
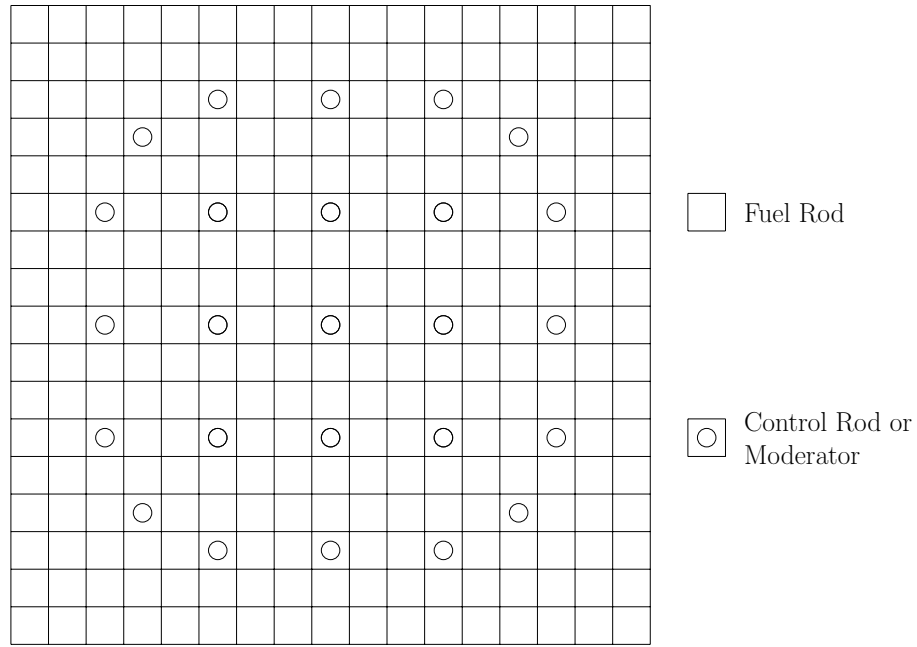

\begin{figure}[H]
\centering
\scalebox{0.5}{\begin{tikzpicture}
\draw[step=0.5,black,thin] (0,0) grid (10,10);
\draw[step=1,black,thin] (11,0) grid (21,10);
\foreach \x in {1.5,2,...,8}
\foreach \y in {4,4.5,...,5.5}
\fill[lightgray] (\x,\y) rectangle ++ (0.5,0.5);
\foreach \x in {1.5,2,...,8}
\foreach \y in {4,4.5,...,5.5}
\fill[lightgray] (\y,\x) rectangle ++ (0.5,0.5);

\foreach \x in {2,2.5,...,7.5}
\foreach \y in {3.5,4,...,6}
\fill[lightgray] (\x,\y) rectangle ++ (0.5,0.5);
\foreach \x in {2,2.5,...,7.5}
\foreach \y in {3.5,4,...,6}
\fill[lightgray] (\y,\x) rectangle ++ (0.5,0.5);

\foreach \x in {2.5,3,...,7}
\foreach \y in {3,3.5,...,6.5}
\fill[lightgray] (\x,\y) rectangle ++ (0.5,0.5);
\foreach \x in {2.5,3,...,7}
\foreach \y in {3,3.5,...,6.5}
\fill[lightgray] (\y,\x) rectangle ++ (0.5,0.5);

\foreach \x in {12,13,...,19}
\foreach \y in {4,5,...,5}
\fill[lightgray] (\x,\y) rectangle ++ (1,1);
\foreach \x in {1,2,...,8}
\foreach \y in {15,16,...,16}
\fill[lightgray] (\y,\x) rectangle ++ (1,1);

\foreach \x in {13,14,...,18}
\foreach \y in {3,4,...,6}
\fill[lightgray] (\x,\y) rectangle ++ (1,1);
\foreach \x in {2,3,...,7}
\foreach \y in {14,15,...,17}
\fill[lightgray] (\y,\x) rectangle ++ (1,1);

\draw[step=1.0,black,thin] (22,6) grid (23,7);
\node[font=\huge,text width = 5cm] at (26,6.5) {Moderator};
\draw[step=1.0,black,thin](22,2) grid (23,3);
\fill[lightgray] (22,2) rectangle ++ (1,1);
\node[font=\huge,text width = 5cm] at (26,2.5) {Moderator, Control Rod or Fuel };
\node[font=\huge,text width = 5cm] at (-5,5.5) {\phantom{Guide-Tube or}};


\end{tikzpicture}}
\caption{Computational grid shown here for a single lattice-cell for the fine ($20 \times 20$) grid (used for the assembly calculations in Section~\ref{sec:fuel_assem_results}) and the coarse ($10 \times 10$) grid (used in the reactor core calculations in Section~\ref{sec:reactor_core_results}).}
\label{fig:fuel_rod_mesh}
\end{figure}
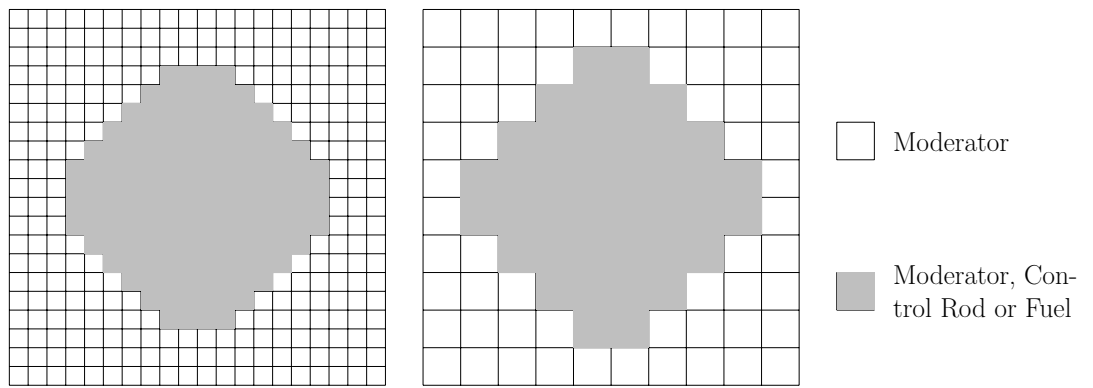

All lattice-cells in the fuel assembly have the same geometry with the moderator occupying the outer region of every lattice-cell and either fuel, a control rod or moderator occupying the inner region. This is shown in Figure~\ref{fig:fuel_rod_mesh}. The guide tube is not modelled. The material parameters required are UOX cross-sections for the fuel rods, and cross-sections for the control rods and moderator (same as the coolant), all taken from the KAIST benchmark.

\subsection{Fuel Assembly - Finite Volume Discretisation}\label{sec:fuel_assem_results}
The neutron diffusion equation is solved for a 2D fuel assembly which uses geometry and cross-sections from the KAIST benchmark~\citep{kaist2000}. We perform two Jacobi iterations, 100 multigrid iterations and 100 multi-group iterations to obtain the solution from the multi-group neural network with weights that are pre-determined by a finite volume discretisation. After the final multi-group iteration, the solution converged to an effective tolerance of $10^{-14}$. Solutions obtained from the neural network are compared with solutions from a traditional implementation of the finite volume discretisation in Fortran that uses a Gauss-Seidel iterative method (with a tolerance of $10^{-15}$). 

\begin{figure}[H]
    \centering
    \includegraphics[scale=1]{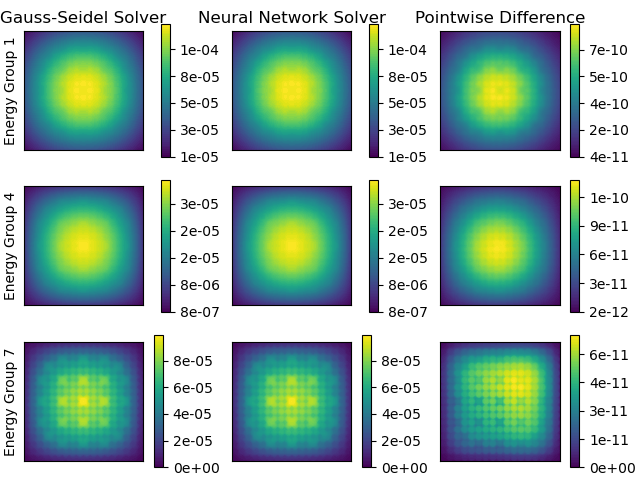}
    \caption{Scalar flux ($\si{neutrons.cm^{-2}.s^{-1}}$) across the fuel assembly for three energy groups for a fuel assembly with control rods fully withdrawn, generated using the multi-group network.}
    \label{fig:fa_no_cr}
\end{figure}

Figure~\ref{fig:fa_no_cr} contains the flux profiles of three energy groups for a fuel assembly with control rods fully withdrawn. The high values of scalar flux for group~7 indicate the location of the moderator within the guide tubes. It can be observed that the pointwise difference between the neural network solution and the Fortran solution with Gauss-Seidel iteration is small, $\mathcal{O}(10^{-10})$, and within the tolerances set for the solvers.

\begin{figure}[H]
    \centering
    \includegraphics[scale=1]{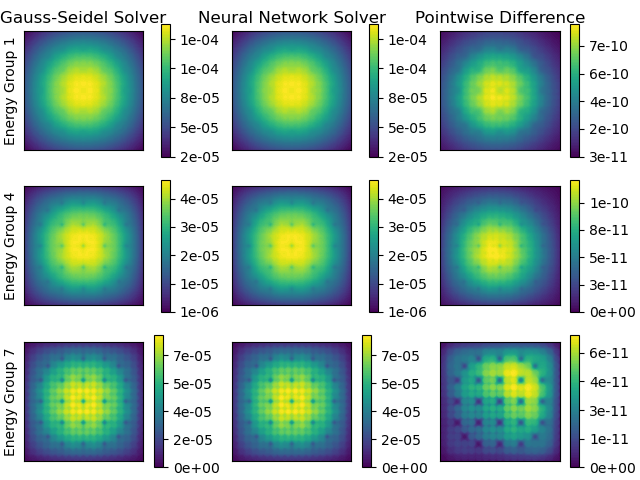}
    \caption{Scalar flux (\si{neutrons.cm^{-2}.s^{-1}}) across the fuel assembly for three energy groups for a fuel assembly with control rods fully inserted, generated using the multi-group network.}
    \label{fig:fa_with_cr}
\end{figure}

Figure~\ref{fig:fa_with_cr} contains the flux profiles for three energy groups for a fuel assembly with control rods fully inserted. The positions of the control rods can be observed in between the fuel rods, where the flux decreases sharply. It can be observed that the pointwise difference between the neural network solution and the Fortran solution is small, $\mathcal{O}(10^{-10})$, and within the tolerances set for the solvers.

\begin{figure}[H]
\centering
\begin{minipage}{.45\textwidth}
  \centering
  \includegraphics[scale=0.5]{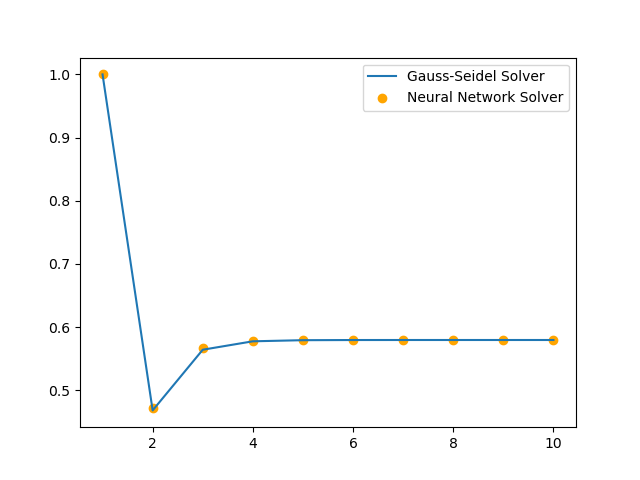}
  \subcaption{Control rods fully withdrawn. $k_{\text{eff}}$ converging to 0.5797 for the Neural Network Solver and 0.5797 for the Fortran implementation (with the Gauss-Seidel solver).}
  \label{fig:fa_keff_no_cr}
\end{minipage}%
\hfill
\begin{minipage}{.45\textwidth}
  \centering
  \includegraphics[scale=0.5]{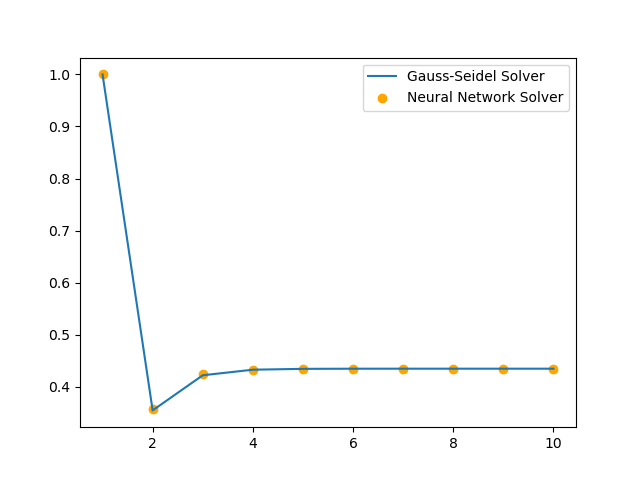}
  \subcaption{Control rods fully inserted. $k_{\text{eff}}$ converging to 0.4347 for the Neural Network Solver and 0.4347 for the Fortran implementation (with the Gauss-Seidel solver).}
  \label{fig:fa_keff_with_cr}
\end{minipage}
\caption{A plot of convergence of $k_{\text{eff}}$ against power iteration for the fuel assembly, generated using the multi-group network with the finite volume discretisation.}
\label{fig:fa_keff}
\end{figure}

Figure~\ref{fig:fa_keff} contains the rate of convergence of $k_{\text{eff}}$ for a fuel assembly with control rods fully withdrawn (Figure~\ref{fig:fa_keff}(a)) and fully inserted (Figure~\ref{fig:fa_keff}(b)). It can be observed that $k_{\text{eff}}$ is lower when control rods are inserted, as would be expected. The convergence for the solution from the neural network solver and that from the Fortran implementation is identical for both configurations (fully withdrawn and fully inserted control rods). See Table~\ref{tab:keff_assembly} for a comparison of the converged values of~$k_{\text{eff}}$.

\subsection{Fuel Assembly - Finite Element discretisation}

\begin{figure}[H]
\centering
\begin{minipage}{.5\textwidth}
  \centering
  \includegraphics[scale=0.55]{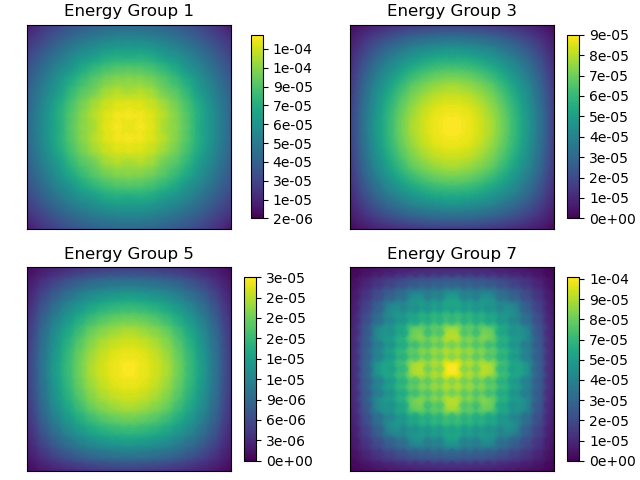}
  \subcaption{Scalar flux (\si{neutrons.cm^{-2}.s^{-1}}) across the fuel assembly for four energy groups.}
  \label{fig:fa_up_out_flux}
\end{minipage}%
\hspace{5mm}
\begin{minipage}{.4\textwidth}
  \centering
  \includegraphics[scale=0.5]{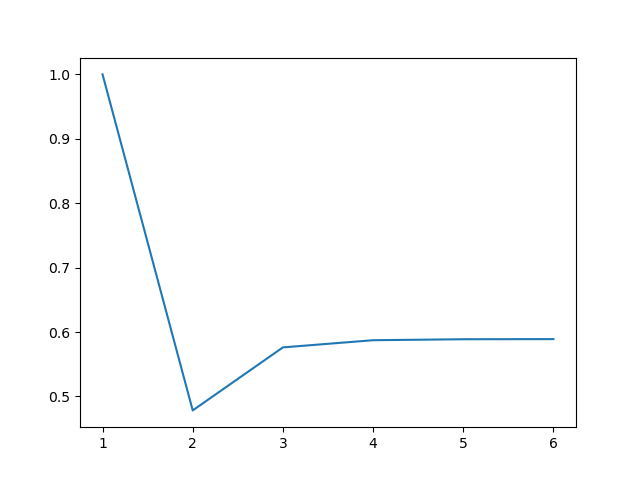}
  \subcaption{Convergence of $k_{\text{eff}}$ against power iteration. $k_{\text{eff}}$ converges to 0.5838 compared to~0.5797 for the finite volume discretisation.}
  \label{fig:fa_up_out_keff}
\end{minipage}
\caption{Results for a fuel assembly with control rods fully withdrawn, generated using the multi-group network for quadratic convolutional finite elements (using ConvFEM).}
\label{fig:fa_up_out_both}
\end{figure}

\begin{figure}[H]
\centering
\begin{minipage}{.5\textwidth}
  \centering
  \includegraphics[scale=0.55]{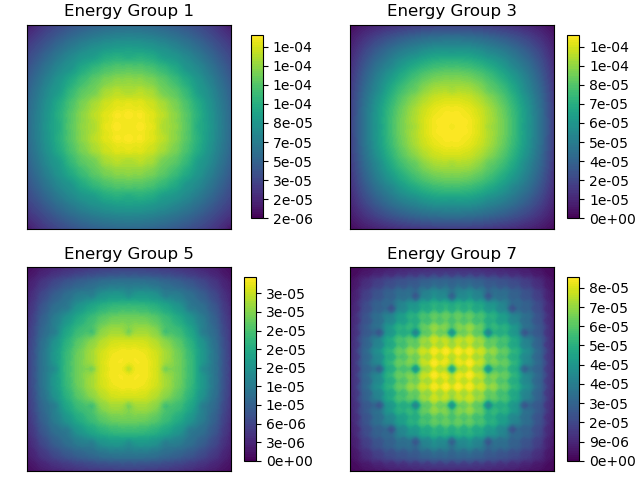}
  \subcaption{Scalar flux (\si{neutrons.cm^{-2}.s^{-1}}) across the fuel assembly for four energy groups.}
  \label{fig:fa_up_in_flux}
\end{minipage}%
\hspace{5mm}
\begin{minipage}{.4\textwidth}
  \centering
  \includegraphics[scale=0.5]{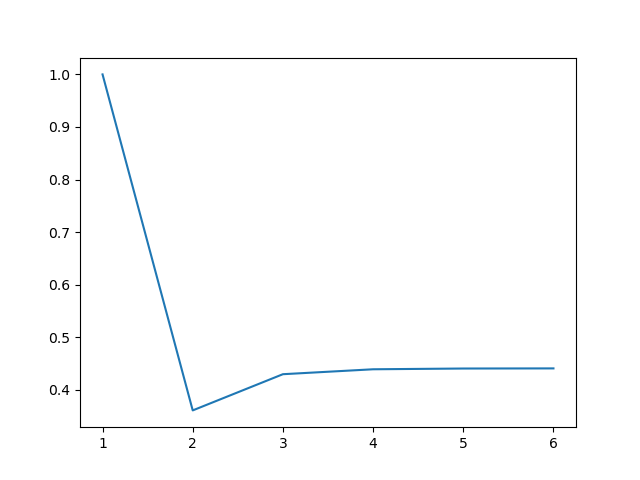}
  \subcaption{$k_{\text{eff}}$ vs power iteration. $k_{\text{eff}}$ converges to 0.4370 compared to 0.4347 for the finite volume discretisation.}
  \label{fig:fa_up_in_keff}
\end{minipage}
\caption{Results for a fuel assembly with control rods fully inserted, generated using the multi-group network for quadratic convolutional finite elements (with ConvFEM).}
\label{fig:fa_up_in_both}
\end{figure}

Figures~\ref{fig:fa_up_out_both} and~\ref{fig:fa_up_in_both} contain the scalar flux solution for a fuel assembly with control rods fully withdrawn and fully inserted, respectively. Both solutions were generated using quadratic convolutional finite elements (ConvFEM) implemented with a neural network. The weights used in the filters are given in Equation~\eqref{eq:quad_filt}. In both cases, the flux profile shows a similar distribution to the solutions generated using the finite volume discretisation in Section~\ref{sec:fuel_assem_results}. The converged values of $k_{\text{eff}}$ using the quadratic finite elements are both slightly larger than for the finite volume discretisation, see Table~\ref{tab:keff_assembly}. 

\begin{table}[htbp]
\centering
\begin{tabular}{lllcc}
\toprule
discretisation         &  implementation   & solver &  withdrawn  & inserted \\
                \toprule
finite volume & neural network &  multigrid with Jacobi iterations & 0.5797       & 0.4347 \\
finite volume & Fortran        &  Gauss-Seidel        & 0.5797       & 0.4347 \\
finite element & neural network &  multigrid with Jacobi iterations & 0.5838      & 0.4370 \\
\bottomrule
\end{tabular}
\caption{Values of $k_{\text{eff}}$ for the fuel assembly test cases}
\label{tab:keff_assembly}
\end{table}

\subsection{Fuel Assembly - Time comparisons}

Table~\ref{tab:fuel_asse_time} shows the time comparisons for 100 Jacobi iterations performed on the fuel assembly test case. The neural network implementation using the GPU used the multi-group network (see figure~\ref{fig:multi_group_net}) but replaced the multigrid network (see figure~\ref{fig:multi_grid_net}) with the Jacobi network (see figure~\ref{fig:jacobi_net}). The equivalent operations were performed in a Fortran code in serial using a CPU. It can be observed that the average time for the neural network solver was less than one-third that of the time for the solver written in Fortran. The neural network solver also shows more consistent timings, with the difference between the minimum and maximum times being 0.0636 seconds. The Fortran solver shows a much greater variance in timings, with the difference between the minimum and maximum times being 1.3400 seconds.

\begin{table}[htbp]
\centering
\begin{tabular}{lllcc}
\toprule
 implementation   & hardware &  max time (s)  & min time (s) & average time (s)\\
                \toprule
 neural network &  NVIDIA RTX 6000 GPU & 1.3412       & 1.2776  & 1.2819  \\
Fortran         &  AMD EPYC 7742 CPU & 5.2568      & 3.9168 &  4.3681\\
\bottomrule
\end{tabular}
\caption{Time comparisons for 100 Jacobi iterations performed on the fuel assembly test case using GPU for the neural network solver and a CPU for the Fortran code. The 100 iterations were performed 400 times so the maximum, minimum and average times from these are shown.}
\label{tab:fuel_asse_time}
\end{table}

\subsection{Reactor Core - Geometry and Configuration}\label{sec:reactor_core_geometry}
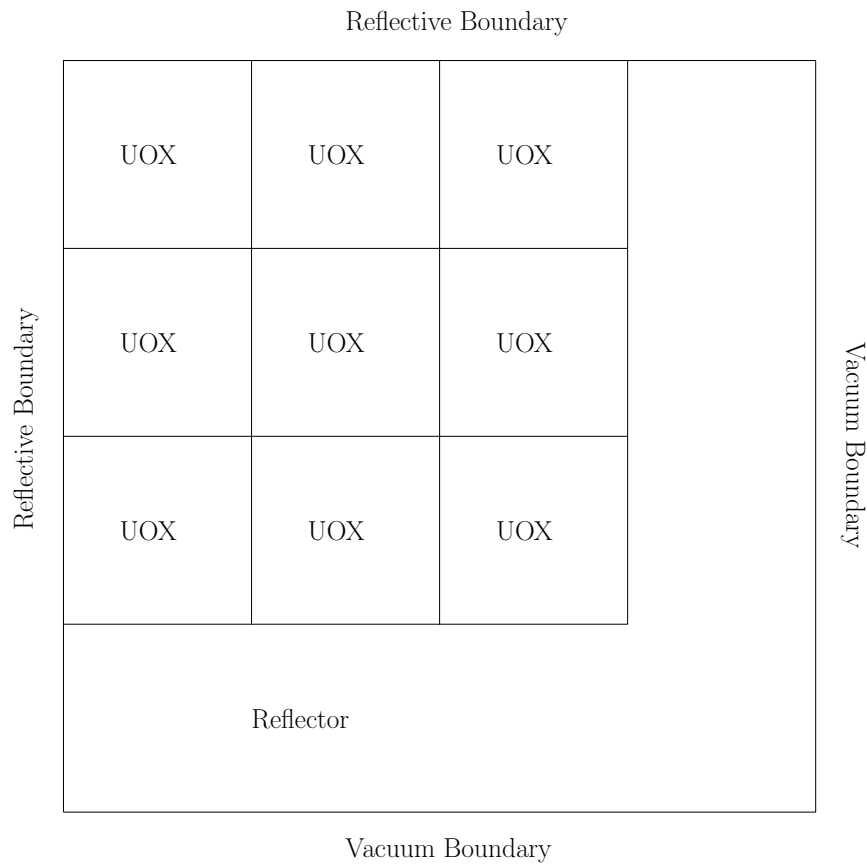
\begin{figure}[H]
\centering
\scalebox{0.5}{\begin{tikzpicture}
\draw[step=20.0,black,thin] (0,0) grid (20,20);
\draw[step=5.0,black,thin] (0,5) grid (15,20);
\foreach \x in {2.5,7.5,12.5}
\foreach \y in {7.5,12.5,17.5}
\node[font=\huge,text width = 2cm] at (\x,\y) {UOX};
\node[font=\huge,text width = 5cm] at (7.5,2.5) {Reflector};

\node[font=\huge,text width = 7cm,rotate=90] at (-1,11) {Reflective Boundary};
\node[font=\huge,text width = 7cm,rotate=-90] at (21,9) {Vacuum Boundary};
\node[font=\huge,text width = 7cm] at (11,-1) {Vacuum Boundary};
\node[font=\huge,text width = 7cm] at (11,21) {Reflective Boundary};



\end{tikzpicture}}
\caption{Geometry of Reactor Core for a simplified version of the KAIST benchmark~\citep{kaist2000}.}
\label{fig:reactor_core_mesh}
\end{figure}

We now model a reactor core using the cross-sections from the KAIST benchmark~\citep{kaist2000}. Unlike the benchmark, our core is a $3\times 3$ grid of fuel assemblies of type UOX only. 
One quarter of the domain is modelled, using reflective boundary conditions to represent the rest of the core, see Figure~\ref{fig:reactor_core_mesh}. The reflector surrounding the fuel assemblies uses the moderator material. Vacuum boundary conditions are applied to the external boundary of the core. The width of the reflector and each of the assemblies is \SI{21.42}{cm} so each side of the domain shown in Figure~\ref{fig:reactor_core_mesh} measures \SI{85.68}{cm}.  Each lattice-cell of the  assemblies has a computational grid of $10 \times 10$ cells (see Figure~\ref{fig:fuel_rod_mesh}). The grid is uniform throughout the domain, meaning that the reflector contains $202,300$ cells, all nine fuel assemblies contain a total of $260,100$ cells and $2,724$ cells are used as halo cells needed to apply the boundary conditions. The energy was again discretised into seven groups resulting in $3,236,800$ degrees of freedom. 

Each fuel assembly can either have control rods fully inserted or fully withdrawn. The two configurations of the core that are investigated here can be seen in Figure~\ref{fig:reactor_config}. 
Configuration one has five fuel assemblies with fully withdrawn control rods and four fuel assemblies with control rods fully inserted. Configuration two has six fuel assemblies with fully withdrawn control rods and three fuel assemblies with fully inserted  control rods. These configurations were chosen randomly.

  \begin{figure}[H]
\centering
\begin{minipage}{.45\textwidth}
  \centering
  \scalebox{0.35}{\begin{tikzpicture}
\draw[step=5.0,black,thin] (0,0) grid (15,15);
\node[font=\huge,text width = 3.5cm] at (2.5,2.5) {Withdrawn};
\node[font=\huge,text width = 3.5cm] at (2.5,7.5) {Withdrawn};
\node[font=\huge,text width = 3.5cm] at (2.5,12.5) {Inserted};
\node[font=\huge,text width = 3.5cm] at (7.5,7.5) {Inserted};
\node[font=\huge,text width = 3.5cm] at (7.5,12.5) {Withdrawn};
\node[font=\huge,text width = 3.5cm] at (7.5,2.5) {Inserted};
\node[font=\huge,text width = 3.5cm] at (12.5,2.5) {Inserted};
\node[font=\huge,text width = 3.5cm] at (12.5,7.5) {Withdrawn};
\node[font=\huge,text width = 3.5cm] at (12.5,12.5) {Withdrawn};




\end{tikzpicture}}
  \subcaption{Reactor configuration one.}
  \label{fig:reactor_config_one}
\end{minipage}%
\hfill
\begin{minipage}{.45\textwidth}
  \centering
  \scalebox{0.35}{\begin{tikzpicture}
\draw[step=5.0,black,thin] (0,0) grid (15,15);
\node[font=\huge,text width = 3.5cm] at (2.5,2.5) {Inserted};
\node[font=\huge,text width = 3.5cm] at (2.5,7.5) {Withdrawn};
\node[font=\huge,text width = 3.5cm] at (2.5,12.5) {Withdrawn};
\node[font=\huge,text width = 3.5cm] at (7.5,7.5) {Inserted};
\node[font=\huge,text width = 3.5cm] at (7.5,12.5) {Withdrawn};
\node[font=\huge,text width = 3.5cm] at (7.5,2.5) {Inserted};
\node[font=\huge,text width = 3.5cm] at (12.5,2.5) {Withdrawn};
\node[font=\huge,text width = 3.5cm] at (12.5,7.5) {Withdrawn};
\node[font=\huge,text width = 3.5cm] at (12.5,12.5) {Withdrawn};




\end{tikzpicture}}
  \subcaption{Reactor configuration two.}
  \label{fig:reactor_config_two}
\end{minipage}
\caption{Reactor core configurations where withdrawn means control rods are fully withdrawn and inserted means control rods are fully inserted.}
\label{fig:reactor_config}
\end{figure}

\subsection{Reactor Core - Finite Volume discretisation}\label{sec:reactor_core_results}
\begin{figure}[H]
    \centering
    \includegraphics[scale=1]{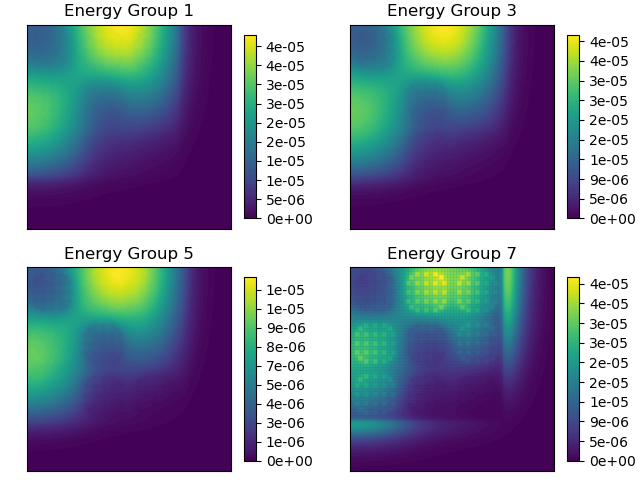}
    \caption{Scalar flux ($\si{neutrons.cm^{-2}.s^{-1}}$) across the fuel assembly for four energy groups for reactor configuration one, generated using the multi-group network with the finite volume discretisation.}
    \label{fig:rc_sol_one}
\end{figure}

A neural network with weights determined by a finite volume discretisation was used to solve the 2D neutron diffusion equation and give solutions for the reactor core described in the previous section. For all the solutions in this section, 5 Jacobi iterations, 100 multigrid iterations and 100 multi-group iterations were performed. Figure~\ref{fig:rc_sol_one} contains the flux profiles for four energy groups for reactor configuration one. It can be observed that flux is higher in regions where control rods are fully withdrawn, with a notable drop  for the flux of all the energy groups in the upper left corner where they are inserted. In the flux profile of the lowest energy group (group~7), the locations of the control rods and the moderator (within the guide tubes) are clearly picked out with the flux decreasing or increasing sharply respectively. 

\begin{figure}[H]
    \centering
    \includegraphics[scale=1]{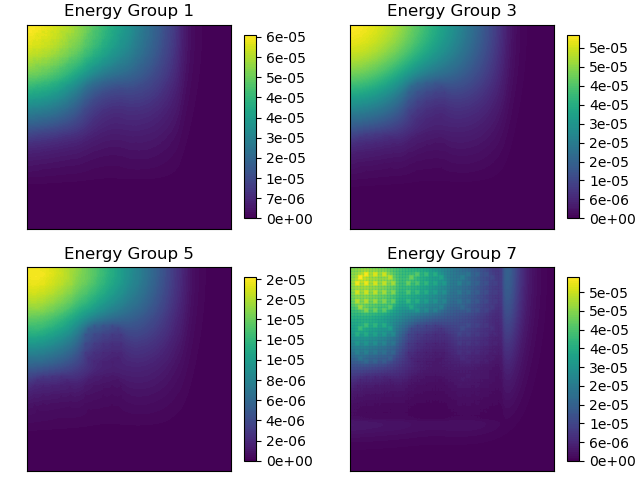}
    \caption{Scalar flux (\si{neutrons.cm^{-2}.s^{-1}}) across the fuel assembly for four energy groups for reactor configuration two, generated using the multi-group network for the finite volume discretisation.}
    \label{fig:rc_sol_two}
\end{figure}

Figure~\ref{fig:rc_sol_two} contains the flux profiles for four energy groups for reactor configuration two. Again, the flux drops sharply  where control rods are inserted, with the highest flux values occurring in the upper left corner by the reflective boundaries. The locations of both the control rods and moderator within the guide tubes are clearly picked out in the flux profile of the lowest energy group (group~7).

\begin{figure}[H]
\centering
\begin{minipage}{.45\textwidth}
  \centering
  \includegraphics[scale=0.5]{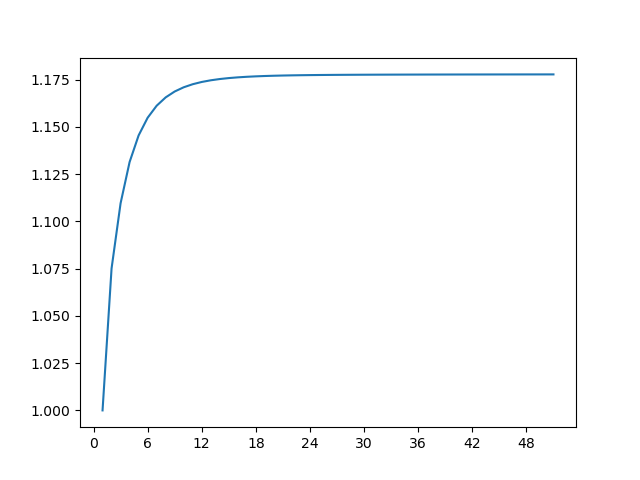}
  \subcaption{$k_{\text{eff}}$ vs power iteration for reactor configuration one, converging to 1.1777.}
  \label{fig:fa_keff_no_cr}
\end{minipage}%
\hfill
\begin{minipage}{.45\textwidth}
  \centering
  \includegraphics[scale=0.5]{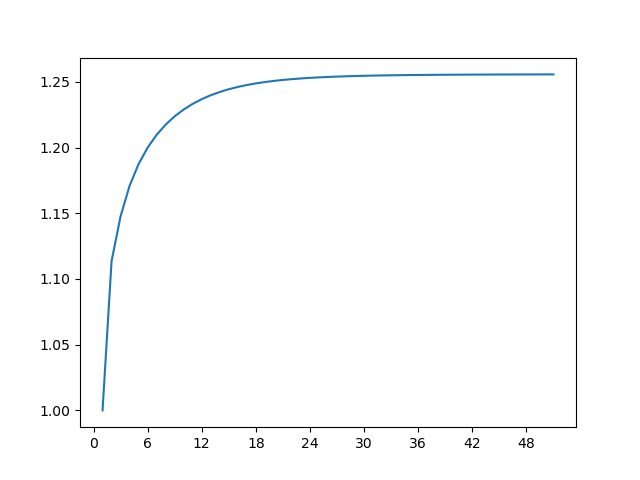}
  \subcaption{$k_{\text{eff}}$ vs power iteration for reactor configuration two,  converging to 1.2557.}
  \label{fig:fa_keff_with_cr}
\end{minipage}
\caption{Reactor core $k_{\text{eff}}$ vs power iteration using the multi-group network with a finite volume discretisation for both configurations.}
\label{fig:rc_keff}
\end{figure}

Figure~\ref{fig:rc_keff} shows the convergence of $k_{\text{eff}}$ for both reactor configurations. It can be observed that configuration two has a slightly higher  $k_{\text{eff}}$ than configuration one, which is expected as configuration two has fewer control rods inserted.

\section{Conclusions and Future work}\label{conclusion}



This paper presents a new approach that uses the tools within Artificial Intelligence (AI) software libraries to replicate the processes of solving partial differential equations that have been discretised through standard numerical method schemes. Whilst applicable to partial differential equations (PDEs) in general, this article has focused on the field of nuclear reactor physics and solves the eigenvalue problem arising from neutron transport, as described through diffusion theory. Furthermore, whilst underlying discretisation methods can be arbitrary, our demonstration focuses on the use of convolutional neural networks to replicate the solution process when using the finite volume method. Instead of training the network, the approach taken here is to define the weights of convolutional neural network in order to reproduce the discretisation exactly. Iterative solvers are also replicated within the network. A sawtooth multigrid method based on the U-Net architecture with an internal Jacobi iteration is investigated here. The multigrid network is then used with another network that acts across all energy groups as a multi-group solver. 

Two test cases are used to demonstrate the approach, a fuel assembly and a reactor core. For the fuel assembly test case, the solution from the neural network solution is compared with the same finite volume discretisation solved by a Gauss-Seidel method and implemented in a standard way using Fortran. The absolute pointwise error between the two solutions was $\mathcal{O}(10^{-10})$. The fuel assembly test case demonstrates that the approach produces the identical solution (accounting for solver tolerances) to that obtained through a standard approach, and produces the same rate of convergence for $k_{\text{eff}}$. This test case is also used to demonstrate how a quadratic finite element discretisation may be used in the convolutional layers. The approach is also extended to a more computationally demanding problem, in the form of a reactor core.

A benefit of using such an approach is that it allows one to exploit the power of AI libraries and their built-in technologies. For example, their executions are already optimised for different computer architectures, whether it be CPUs, GPUs or new-generation AI processors. This flexibility brings within easy reach the ability to run code on multiple platforms without the need for modification of the code. A further benefit is that of simplified code development, as the AI libraries abstract away code relating to the platform, leaving the user to concentrate on their programming tasks. As well as exploiting the substantial developments already made in AI libraries, formulating numerical discretisations as convolutional layers in neural networks will mean that these codes are ready to run on the latest AI processors.

Future work will involve including the power eigenvalue iteration within the neural network. This would enable the neural network to calculate sensitivities of the eigenvalue to material properties automatically, using the backpropagation algorithm of the neural network. An important next step would be to optimise the code and methods further (e.g.~taking into account the multigrid bottleneck caused by the coarsest grid) so that large problems can be run on GPUs or new AI computers.

\section*{CRediT authorship contribution statement}

\textbf{TRFP:} methodology, software, writing (original draft, review and editing).
\textbf{CEH:} methodology, writing (original draft, review and editing), supervision.
\textbf{BC:} software, writing (review and editing).
\textbf{AGB:} software, writing (original draft,  review and editing).
\textbf{CCP:} conceptualisation, methodology, software, writing (original draft, review and editing), supervision, funding acquisition.

\section*{Acknowledgements}
The authors would like to acknowledge the following EPSRC grants: RELIANT, Risk EvaLuatIon fAst iNtelligent Tool for COVID19 (EP/V036777/1); CO-TRACE, COvid-19 Transmission Risk Assessment Case Studies --- education Establishments (EP/W001411/1); INHALE,
Health assessment across biological length scales (EP/T003189/1); the PREMIERE programme grant (EP/T000414/1); MAGIC (EP/N010221/1); and MUFFINS (EP/P033180/1).

\bibliography{mybibfile}

\appendix
\section{Diffusion operator}\label{sec:diffusion_operator_rewritten}
For two scalars $\phi_g$ and $D_g$, the following is true
\begin{eqnarray}
\nabla^2(D_g\phi_g) & = & \nabla\cdot\nabla(D_g\phi_g) \\
                    & = & \nabla\cdot \left( \phi_g\nabla D_g + D_g\nabla \phi_g\right)\\
                    & = &  \phi_g\nabla^2 D_g + D_g\nabla^2 \phi_g+ 2\nabla D_g \cdot \nabla\phi_g\,, 
\end{eqnarray}
which leads to the following identity
\begin{equation}
\label{eq:differential_identity}
2\nabla D_g \cdot \nabla\phi_g = \nabla^2(D_g\phi_g) - \phi_g\nabla^2 D_g - D_g\nabla^2 \phi_g\,.
\end{equation}
Expanding out the diffusion term in Equation~\eqref{eq:diff-eig} and then substituting in the expression from Equation~\eqref{eq:differential_identity} results in
\begin{eqnarray}
\nabla\cdot(D_g\nabla\phi_g) & = &  D_g \nabla^2\phi_g + \nabla\phi_g \cdot \nabla D_g\\
 & = &  D_g \nabla^2\phi_g + \frac{1}{2}\left( \nabla^2(D_g\phi_g) - \phi_g\nabla^2 D_g - D_g\nabla^2 \phi_g\right)\\
  & = & \frac{1}{2}\left( \nabla^2(D_g\phi_g) - \phi_g\nabla^2 D_g + D_g\nabla^2 \phi_g\right) \,.\label{eq:final_form}
\end{eqnarray}
Equation~\eqref{eq:final_form} is used in Equation~\eqref{eq:analytical_form}.

\section{Equivalence of finite volume discretisation written in standard notation and written as convolutions}\label{sec:equivalence}

First, let us recall that the Hadamard product of two $N$ by $M$ matrices is given by
\begin{equation}
\bm{A}\odot \bm{B}\big|_{k\ell} = A_{k\ell}B_{k\ell}\quad \forall k\in\{1,2,\dots,N\},\,\ell\in\{1,2,\dots,M\}
\end{equation}
and the sign $\displaystyle{\sum_{\text{entries}}}$ sums all the entries of a matrix
\begin{equation}\label{eq:definition_sum_entries}
\sum_{\text{entries}} \bm{A} \equiv \sum_{k=1}^{M} \sum_{\ell=1}^{N} A_{k\ell}\,.
\end{equation}

In this section, we will show equivalence of the diffusion operator's finite volume discretisation given in Equation~\eqref{eq:discretised_hfm} (also in  Equations~\eqref{eq:discretised_hfm_rewritten} and~\eqref{eq:define_aijuv}) and the same discretisation formulated as a convolutional layer with pre-defined weights as described by Equations~\eqref{eq:Laplacian_convolution} and~\eqref{eq:discretised_form}. Considering each term on the right-hand side of Equation~\eqref{eq:discretised_form}, we start with part of the second term and evaluating this in the $i,j$th cell: 
\begin{eqnarray}
\bm{f}(\bm{\Phi_g};\bm{w})\Big|_{i,\,j} & = & \sum_{\text{entries}}  \begin{bmatrix}
 0
 & \frac{-1\phantom{-}}{\Delta y^2} & 0 \\
\frac{-1\phantom{-}}{\Delta x^2} & \frac{2}{\Delta x^2} +\frac{2}{\Delta y^2} &\frac{-1\phantom{-}}{\Delta x^2} \\
 0 & \frac{-1\phantom{-}}{\Delta y^2} & 0
 \end{bmatrix}\odot 
 \begin{bmatrix*}[l]
 \phi_{i-1,j+1,g} & \phi_{i,j+1,g} & \phi_{i+1,j+1,g} \\
 \phi_{i-1,j,g} & \phi_{i,j,g} & \phi_{i+1,j,g} \\
 \phi_{i-1,j-1,g} & \phi_{i,j-1,g} & \phi_{i+1,j-1,g} \\
 \end{bmatrix*} \\[3mm]
& = & \frac{-\left( \phi_{i-1,j,g} + \phi_{i+1,j,g}\right)}{\Delta x^2} + \frac{-\left( \phi_{i,j-1,g} + \phi_{i,j+1,g}\right)}{\Delta y^2}+ \left(\frac{2}{\Delta x^2} + \frac{2}{\Delta y^2}\right)\phi_{i,j,g}\,.\quad 
\end{eqnarray}
Now, considering the second term in its entirety,
\begin{multline}\label{eq:second_term}
\left(\bm{D_g}\odot\bm{f}(\bm{\Phi_g};\bm{w})\right)\Big|_{i,\,j}  =  \frac{-D_{i,j,g}\left( \phi_{i-1,j,g} + \phi_{i+1,j,g}\right)}{\Delta x^2} + \frac{-D_{i,j,g}\left( \phi_{i,j-1,g} + \phi_{i,j+1,g}\right)}{\Delta y^2} \\ + \left(\frac{2}{\Delta x^2} + \frac{2}{\Delta y^2}\right)D_{i,j,g}\phi_{i,j,g}\,.\quad
\end{multline}

Similarly, for the third term on the right-hand side of Equation~\eqref{eq:discretised_form}
\begin{multline}\label{eq:third_term}
\left(\bm{\Phi_g}\odot\bm{f}(\bm{D_g};\bm{w})\right)\Big|_{i,\,j}  =  \frac{-\phi_{i,j,g}\left( D_{i-1,j,g} + D_{i+1,j,g}\right)}{\Delta x^2} + \frac{-\phi_{i,j,g}\left( D_{i,j-1,g} + D_{i,j+1,g}\right)}{\Delta y^2} \\ + \left(\frac{2}{\Delta x^2} + \frac{2}{\Delta y^2}\right)D_{i,j,g}\phi_{i,j,g}\,.\quad
\end{multline}

The first term on the right-hand side of Equation~\eqref{eq:discretised_form} can be expanded as follows
\begin{multline}\label{eq:first_term}
\bm{f}\left(\bm{D_g}\odot\bm{\Phi_g};\bm{w}\right)\Big|_{i,\,j}  =  \frac{-\left( D_{i-1,j,g} \phi_{i-1,j,g} + D_{i+1,j,g}\phi_{i+1,j,g}\right)}{\Delta x^2} + \frac{-\left( D_{i,j-1,g} \phi_{i,j-1,g} + D_{i,j+1,g}\phi_{i,j+1,g}\right)}{\Delta y^2} \\ + \left(\frac{2}{\Delta x^2} + \frac{2}{\Delta y^2}\right)D_{i,j,g}\phi_{i,j,g}\,.\quad
\end{multline}
Combining the expressions in Equations~\eqref{eq:second_term}, \eqref{eq:third_term} and \eqref{eq:first_term} according to the definition of the diffusion operator from Equation~\eqref{eq:discretised_form} and gathering terms that multiply each scalar flux term gives
\begin{eqnarray}
\bm{f}^{\text{Diff}}(\bm{\Phi}_g,\bm{D}_g; \bm{w}) &= &\frac{1}{2}\left( \bm{f}(\bm{D}_g\odot\bm{\Phi}_g;\bm{w}) + \bm{D}_g\odot \bm{f}(\bm{\Phi}_g;\bm{w}) - \bm{\Phi}_g\odot \bm{f}(\bm{D}_g;\bm{w})\right) \\[4mm]
& = &  
\frac{-\left( D_{i-1,j,g} \phi_{i-1,j,g} + D_{i+1,j,g}\phi_{i+1,j,g}\right)}{2\Delta x^2} + \frac{-\left( D_{i,j-1,g} \phi_{i,j-1,g} + D_{i,j+1,g}\phi_{i,j+1,g}\right)}{2\Delta y^2} \nonumber \\
&& \;+\; \left(\frac{1}{\Delta x^2} + \frac{1}{\Delta y^2}\right)D_{i,j,g}\phi_{i,j,g}\\[2mm]
&& \;+\; \frac{-D_{i,j,g}\left( \phi_{i-1,j,g} + \phi_{i+1,j,g}\right)}{2\Delta x^2} + \frac{-D_{i,j,g}\left( \phi_{i,j-1,g} + \phi_{i,j+1,g}\right)}{2\Delta y^2} + \left(\frac{1}{\Delta x^2} + \frac{1}{\Delta y^2}\right)D_{i,j,g}\phi_{i,j,g}\nonumber\\
&& \;-\; \frac{-\phi_{i,j,g}\left( D_{i-1,j,g} + D_{i+1,j,g}\right)}{2\Delta x^2} - \frac{-\phi_{i,j,g}\left( D_{i,j-1,g} + D_{i,j+1,g}\right)}{2\Delta y^2}  - \left(\frac{1}{\Delta x^2} + \frac{1}{\Delta y^2}\right)D_{i,j,g}\phi_{i,j,g} \nonumber\\[3mm]
& = &
- \left( \frac{D_{i-1,j,g} + D_{i,j,g}}{2\Delta x^2}\right) \phi_{i-1,j,g} - 
\left( \frac{D_{i,j,g} + D_{i+1,j,g} }{2\Delta x^2}\right) \phi_{i+1,j,g} \nonumber
\\
&& - 
\left( \frac{D_{i,j-1,g} + D_{i,j,g}}{2\Delta y^2} \right)\phi_{i,j-1,g} - 
\left( \frac{D_{i,j,g} + D_{i,j+1,g}}{2\Delta y^2}\right) \phi_{i,j+1,g} \label{eq:equivalence}
\\
&& \;+\; \left( \frac{D_{i-1,j,g} + 2D_{i,j,g} + D_{i+1,j,g}}{2\Delta x^2} + \frac{D_{i,j-1,g} + 2D_{i,j,g} + D_{i,j+1,g}}{2\Delta y^2}  \right) \phi_{i,j,g} \,.\nonumber
\end{eqnarray}
From this, we can see that Equation~\eqref{eq:equivalence} is equivalent to the discretised diffusion operator seen in Equation~\eqref{eq:discretised_hfm}. In other words, this particular finite volume discretisation can be written as a convolutional layer in a neural network with a 3~by~3 kernel or filter with weights
\begin{equation}
\bm{w} = \begin{bmatrix}
 0
 & \frac{-1\phantom{-}}{\Delta y^2} & 0 \\
\frac{-1\phantom{-}}{\Delta x^2} & \frac{2}{\Delta x^2} +\frac{2}{\Delta y^2} &\frac{-1\phantom{-}}{\Delta x^2} \\
 0 & \frac{-1\phantom{-}}{\Delta y^2} & 0
 \end{bmatrix}
\end{equation}
\end{document}